\documentclass[%
               a4paper,
               intlimits,
               ]
               {JHEP3}

\usepackage{graphics}
\usepackage{graphicx}
\usepackage{epsfig}
\usepackage[dvips]{}
\usepackage{axodraw}

\usepackage{colordvi}

\usepackage{latexsym}
\usepackage{amssymb}
\usepackage{amsmath}

\renewcommand{\rho}{\varrho}
\renewcommand{\phi}{\varphi}
\renewcommand{\epsilon}{\varepsilon}
\renewcommand{\theta}{\vartheta}

  \newcommand{\fract}[2]{{\textstyle{\frac{#1}{#2}}}}
  \newcommand{\fracT}[2]{{\genfrac{}{}{}{0}{#1}{#2}}}

  \newcommand{\LL}{\left}
  \newcommand{\RR}{\right}

  \newcommand{\ssh}{\,\slash\!\!\!}
  \newcommand{\SSH}{\,\slash\!\!\!\!}

  \newcommand{\dd}{{\rm d}}
  \newcommand{\Id}{\mathbb{I}}

\newcommand{\ScaledGlueLoop}[4]{
   { \SetScale{#1}
  \SetScaledOffset(#2,#3)
  \begin{picture}(130,80)(-65,-40)
    \ArrowLine(-60,-35)(-50,0)
    \ArrowLine(-50,0)(-60,35)
    \Gluon(-50,0)(-20,0){5.0}{3}
    \Vertex(-20,0){1}
    \GlueArc(0,0)(20,0,180){5.0}{5}
    \GlueArc(0,0)(20,180,360){5.0}{5}
    \Vertex(20,0){1}
    \Gluon(20,0)(50,0){5.0}{3}
    \ArrowLine(60,-35)(50,0)
    \ArrowLine(50,0)(60,35)
    \Text(-1,-45)[]{#4}
  \end{picture}}}
\newcommand{\ScaledTadpole}[4]{
   {
   \SetScale{#1}
  \SetScaledOffset(#2,#3)
  \begin{picture}(130,80)(-65,-40)
    \ArrowLine(-60,-35)(-50,0)
    \ArrowLine(-50,0)(-60,35)
    \Gluon(-50,0)(50,0){5.0}{6}
    \GlueArc(0,21)(15,-90,270){5.0}{6}
    \Vertex(0,4){1}
    \ArrowLine(60,-35)(50,0)
    \ArrowLine(50,0)(60,35)
     \Text(-1,-45)[]{#4}
  \end{picture}}}

\newcommand{\ScaledGhostLoop}[4]{
   {\SetScale{#1}
  \SetScaledOffset(#2,#3)
  \begin{picture}(130,80)(-65,-40)
    \ArrowLine(-60,-35)(-50,0)
    \ArrowLine(-50,0)(-60,35)
    \Gluon(-50,0)(-20,0){5.0}{3}
    \Vertex(-20,0){1}
    \DashArrowArc(0,0)(20,0,180){5}
    \DashArrowArc(0,0)(20,180,360){5}
    \Vertex(20,0){1}
    \Gluon(20,0)(50,0){5.0}{3}
    \ArrowLine(60,-35)(50,0)
    \ArrowLine(50,0)(60,35)
    \Text(-1,-45)[]{#4}
  \end{picture}}}

\newcommand{\ScaledMajoranaLoop}[4]{
   {\SetScale{#1}
  \SetScaledOffset(#2,#3)
  \begin{picture}(130,80)(-65,-40)
    \ArrowLine(-60,-35)(-50,0)
    \ArrowLine(-50,0)(-60,35)
    \Gluon(-50,0)(-20,0){5.0}{3}
    \Vertex(-20,0){1}
    \ArrowArc(0,0)(20,0,180)
    \ArrowArc(0,0)(20,180,360)
    \Vertex(20,0){1}
    \Gluon(20,0)(50,0){5.0}{3}
    \ArrowLine(60,-35)(50,0)
    \ArrowLine(50,0)(60,35)
    \Text(-1,-45)[]{#4}
  \end{picture}}}

\newcommand{\ScaledMajoranaSelfEnergy}[4]{
   {\SetScale{#1}
  \SetScaledOffset(#2,#3)
  \begin{picture}(130,80)(-65,-40)
    \ArrowLine(-60,0)(-20,0)
    \Vertex(-20,0){1}
    \GlueArc(0,0)(20,0,180){5.0}{6}
    \ArrowLine(-20,0)(20,0)
    \Vertex(20,0){1}
    \ArrowLine(20,0)(60,0)
    \Text(-1,-45)[]{#4}
  \end{picture}}}

\title{Non-Commutative (Softly Broken) Supersymmetric Yang--Mills--Chern--Simons}

\author{Nicola Caporaso$^{a,b}$ and Sara Pasquetti$^{c}$
        \\
    ${}^a$:
    Center for Theoretical Physics, MIT, Cambridge, MA 02139 USA
    \\
    \\
    ${}^b$:
    Dipartimento di Fisica, Polo Scientifico, Universit\`a di Firenze; INFN Sezione di Firenze
    ---
    Via  G. Sansone 1, 50019 Sesto Fiorentino, Italy
    \\

    \\
    ${}^c$:
    Dipartimento di  Fisica, Universit\`a  di Parma; INFN--Gruppo Collegato di Parma
    ---
    Parco Area delle Scienze 7/A, 43100 Parma, Italy
    \\
    \\
    E-mails:
    \email{caporaso@fi.infn.it},
    \email{pasquetti@fis.unipr.it},
    }

\abstract {we study $d=2+1$ non-commutative U(1) YMCS,
concentrating on the one-loop corrections to the propagator
and to the dispersion relations.
Unlike its commutative counterpart, this model
presents divergences and hence an IR/UV mechanism,
which we regularize by adding a Majorana gaugino of mass $m_f$,
that provides (softly broken) supersymmetry.
The perturbative vacuum becomes stable for a wide range of coupling
and mass values, and tachyonic modes are generated only in two
regions of the parameters space. One such region corresponds to
removing the supersymmetric regulator ($m_f \gg m_g$), restoring the
well-known IR/UV mixing phenomenon. The other one  (for $m_f \approx
m_g/2$ and large $\theta$) is novel and peculiar of this model.
The two tachyonic regions turn out to be very different in nature.
We conclude with some remarks on the theory's off-shell unitarity.

}

\preprint{\today}

\begin{document}

\section{Introduction}

Non-commutative quantum field theory is a fascinating theoretical
laboratory where non-trivial deformations of spacetime structures
induce novel and unexpected dynamical effects at the quantum
level. Recently they have attracted a lot of attention, mainly due
to the discovery of their relation to string/M theory
\cite{Connes:1997cr,Seiberg:1999vs}. In particular, Seiberg and
Witten realized~\cite{Seiberg:1999vs} that a certain class of
quantum field theories on non-commutative Minkowski spacetimes can
be obtained as a particular low-energy limit of open strings in
the presence of a constant NS-NS $B$-field. From a purely
field-theoretical point of view, they appear as a peculiar
non-local deformation of conventional quantum field theory,
presenting a large variety of new phenomena, not completely
understood even at the perturbative level.
For a self-contained review of these topics, see \cite{Douglas:2001ba,Szabo:2001kg}.

Four-dimensional non-commutative gauge theories are affected by
the infamous IR/UV mixing~\cite{Minwalla:1999px,Matusis:2000jf}
that complicates the renormalization program and may produce
tachyonic instabilities~\cite{VanRaamsdonk:2001jd,Armoni:2001uw}.
Three-dimensional topologically massive electrodynamics (YMCS) is
particularly interesting to study in this respect, for at least two reasons.
First of all, the presence of a single physical
polarization and of an explicit gauge-boson mass
simplifies the analysis of the IR/UV mixing, and
elucidates the nature of the tachyonic instabilities.
In \cite{Caporaso:2004hq}, we have presented some preliminary results,
showing that the purely gauge model suffers from the IR/UV phenomenon,
and that (softly broken) supersymmetry can be employed to remove the IR
singularity at one loop.
In \cite{Asano:2004hy,Ferrari:2005kx} the IR/UV mixing was studied in the exactly supersymmetric theory
employing the superfield formalism, and considering gauge superpotentials
in the fundamental and adjoint representations.
A second reason for our interest in this model is that
non-commutative gauge theories with Chern--Simons terms have been
proposed as effective description of the Fractional Quantum Hall Effect
\cite{Susskind:2001fb,Polychronakos:2001mi,Barbon:2001dw,Cappelli:2004xk}.

Naively one may expect that there is no problem with the IR/UV
mixing for the YMCS system. In fact, topologically massive
commutative  gauge theories are super-renormalizable models, that
actually turn out to be UV-finite in perturbation theory. Thus,
apparently, there is no UV divergence to be moved in the IR region.
However, the finiteness of these theories originates  from their
symmetries: the simultaneous presence of Lorentz- and
gauge-invariance prevents the potential linear divergences. We will
show that, in the non-commutative setup, the linear divergences will
reappear in the infrared through the IR/UV phenomenon because of the
loss of Lorentz invariance. However, the theory is still UV-finite,
and the planar sector won't need any explicit regularization. The
infrared divergences can instead be regulated by introducing softly
broken supersymmetry, through the coupling with a Majorana fermion
\cite{Caporaso:2004hq,Alvarez-Gaume:2003mb,Ferrari:2004pe}.

Our main results concern the analysis of the renormalization of the polarization tensor at one loop.
We concentrate on the dispersion relations $E(\vec p)$,
which are non-trivial because of the loss of Lorentz invariance, like in the
case of theories at finite temperature.
We observe several exotic phenomena,
like a $2+1$-dimensional relic of birefringence, and anomalous dispersion relations.
Thanks to soft supersymmetry and to the presence of a topological mass term,
this model has a perturbatively stable vacuum for a wide range of couplings and masses.
Two tachyonic regions are present, however.
The first one corresponds to the limit in which fermions are removed from the spectrum,
so that supersymmetry is lost and the IR/UV phenomenon is restored.
The second tachyonic region is truly novel, however.
It is generated when the gauge boson's mass $m_g$ is just below the threshold for the decay
in two fermions, $m_g\lesssim 2 m_f$, and for strong non-commutativity.
The reason for its occurence is related to the fact that, in a non-commutative theory,
the shift in the mass renormalization $m^{\rm R}-m^{\rm bare}$ of a particle
is a physically observable quantity, and depends on the physical cut-off $\Lambda_{\rm eff}=(\theta p)^{-1}$.

The structure of the paper is as follows. In
section~\ref{sec:Preliminaries} we introduce the
Yang--Mills--Chern--Simons model and its supersymmetric extension,
and describe their symmetries.
Section~\ref{sec:radiative_corrections} is devoted to an analysis
of the properties of the polarization tensor of the theory: in
particular, we show that $\Pi_{\mu\nu}$ is transverse at one loop
using the BRST identities.
Next, we compute the one-loop corrections
to the propagators of the gauge boson and of the gaugino, and
their renormalization.

Sections~\ref{sec:dispersion_relations} and \ref{sec:why_they_occur}
are devoted to a study of the dispersion relations
in several different limits.
We discuss the generation of anomalous dispersion relations and tachyonic instabilities
and the physical mechanisms at work in the two destabilization regions.
In section \ref{sec:IRstuff} we substantiate
some of our claims concerning the consistency of our results.
We check the validity of our findings against the contributions of higher orders in perturbation theory.
We discuss the extension of our analysis of the theory's unitarity off-shell, and
provide an independent check that the tachyonic instabilities encountered in
the previous sections are not due to gauge artifacts.
In section \ref{sec:conclusions} we speculate on the nature of the new vacuum.
An appendix is devoted to establishing the notations, and to the Feynman rules.

\section{Non-commutative SUSY U$(1)$ Yang--Mills--Chern--Simons}\label{sec:Preliminaries}

In this section we introduce our model, a non-commutative U$(1)_\star$
Yang--Mills--Chern--Simons (YMCS) theory with softly broken supersymmetry,
and we discuss its symmetries and main dymanical properties.

The non-commutative
U$(1)_\star$ YMCS Lagrangian is
formally obtained by substituting pointwise products with
Moyal products\footnote{\label{foot:3dim}
  In what follows the case where the non-commutativity tensor $\theta_{\mu\nu}$ in
\[
  [x_{\mu},x_\nu]=i\theta_{\mu\nu}
\]
  is spacelike $(\theta^{0\mu}=0 )$ is intended
  unless specified otherwise. Therefore, $\theta^{\mu\nu}$ is
  a shorthand for $\theta\epsilon^{0\mu\nu}$;
  $\rho$ and $\sigma$ are coordinates on the non-commutativity plane, and
  by $(\theta^{-1})_{\mu\nu}$ we mean the inverse of $\theta_{\mu\nu}$ restricted
  to this plane.
  }
\begin{align}
  \label{MoyalDef}
    \LL(f\star g\RR)(x)
    \doteq
    \int
    \frac{\dd y_{\rho}\,\dd y_{\sigma}}{\pi\vartheta}
    \int
    \frac{\dd z_{\rho}\,\dd z_{\sigma}}{\pi\vartheta}
    f(y)g(z)
    e^{-2i\LL(\vartheta^{-1}\RR)_{\mu\nu}(y-x)^\mu(z-x)^\nu}
\end{align}
inside the commutative U$(N)$ Lagrangian, and then setting $N=1$.
For reasons which will become clear as we proceed, we will be considering
the addition of matter to the model:
\begin{align} \label{symcs_action}
  \begin{split}
    S^{\rm NC}_\text{S-YMCS}=
    \int \dd^3x
    \,
    \Bigg(&
    -\frac{1}{4}F_{\mu\nu}\star F^{\mu\nu}
    +\frac{1}{2}\bar \lambda \star \LL(i\SSH D-m_f\RR)\star\lambda
    +\\&
    -\frac{1}{2}m_g\varepsilon^{\lambda\mu\nu}A_{\lambda}\star\partial_{\mu}A_\nu
    +\frac{i}{3}gm_g\varepsilon^{\lambda\mu\nu}A_{\lambda}\star A_{\mu}\star A_\nu
    \Bigg)
    .
  \end{split}
\end{align}
The first line contains the non-commutative Yang--Mills action and
a minimally coupled Majorana fermion; the Chern--Simons term
is in the second line.

There are a few subtleties involved in the definition of this action, related to
the impossibility of employing general gauge groups, and of
implementing local observables~\cite{Szabo:2001kg}.
The action~\eqref{symcs_action}  is invariant under
star-gauge transformations
\begin{align}
    A_\mu^u(x)
    \doteq
    u(x)\star A_\mu(x)\star u^\dag(x)
    -\frac{i}{g}\partial_\mu u(x) \star u^\dag(x),
\end{align}
generated by star-unitary functions: $u(x)\star u^\dag(x) = u(x)^\dag\star u(x)= 1$. Infinitesimally, for $u\doteq 1-ig\epsilon+{\rm O}(g^2)$,
\begin{align}
    {\cal D}_\mu\epsilon
    \doteq
    A_\mu(x)-A^\epsilon_\mu(x)
    =
    \partial_\mu\epsilon-ig[A_\mu,\epsilon]_\star
     .
\end{align}
The field strength $F_{\mu\nu}\doteq \partial_\mu A_\nu -
\partial_\nu A_\mu -ig[A_\mu,A_\nu]_\star$
is in the adjoint representation: $F^u(x)=u(x)\star
F_{\mu\nu(x)}\star u^\dag(x)$, which shows that the Yang--Mills
term $\int \dd^3x \,F_{\mu\nu} \star F^{\mu\nu}$ is star-gauge invariant.

Unlike its commutative counterpart, a U$(N)_\star$ theory is
interacting also in the $N=1$ case, owing to the fact
that star-commutators do not vanish:
\begin{align}
    [A_\mu (x),A_\nu(x)]_\star
    =
    2i A_\mu(x)
    \sin\LL(
    \frac{\stackrel{\leftarrow}{\partial_\rho}
          \theta^{\rho\sigma}
      \stackrel{\rightarrow}{\partial_\sigma}}
      {2}
      \RR)
    A_\nu(x)
    .
\end{align}
The group of star-unitary transformations is vast, as it includes
translations, rotations and area-preserving diffeomorphisms. This
allows for the non-trivial structure constants $\sin(\frac{k\theta
p}{2})$, which can be obtained as a specific $N\rightarrow\infty$
limit of the commutative U$(N)$ structure constants.
The U$(1)_\star$ case therefore captures most of the
perturbative dynamics of the U$(N)_\star$ theories, where the
colour structure constants $f^{abc}$ factorize in front of all
Feynman diagrams.

Taking the $\theta\rightarrow 0$ limit, all interactions
are turned off and one retrieves the ordinary, free U$(1)$ theory.
As this limit may be ill-defined because of infrared divergences,
we choose to work with a finite value of $\theta$,
and expand only in the dimensionless coupling constant $g^2/m_g$,
obtaining a perturbative dynamics that
is closely related to that of ordinary U$(N)$ theories, with three-
and four-gauge boson vertices.
In addition, we remark that truncating the Moyal product
\eqref{MoyalDef} at any finite order in $\theta$ entails losing
all information about the non-locality of the theory.

The Chern--Simons piece in the second line of~\eqref{symcs_action} provides a topological mass for the
gauge boson \cite{Deser:1981wh,Deser:1982vy}, and is not invariant under
star-gauge transformations: its shift is given by
\begin{align}
  \begin{split}
    \frac{m_g}{6g^2}
    \int \dd^3x \,\epsilon_{\lambda\mu\nu}
    (
     u^\dag \star \partial_\lambda u \star
     u^\dag \star \partial_\mu u \star
     u^\dag \star \partial_\nu u
    )
    -\frac{m_g}{2g}
    \int \dd^3x \,\epsilon_{\lambda\mu\nu}
    \partial_\lambda (u^\dag\star\partial_\mu u \star A_\nu)
    \doteq\\\doteq
    4\pi^2\LL(\frac{m_g}{g^2}\RR) w(u)
    +\text{total divergence}
  \end{split}
\end{align}
where $w(u)$ is a ``topological" term.

Non-commutative field theories support a variety of classical solutions and
topologically non-trivial field configurations~\cite{Gross:2000ss,Nair:2001rt,Lechtenfeld:2001aw,Lechtenfeld:2001uq}.
One may explicitly construct a transformation which has a
nonvanishing index $w(u)$, in terms of a star-projector
($\mathbb{P}\star \mathbb{P}=\mathbb{P}$) which depends only on the
non-commuting coordinates $\mathbb{P}(x_\rho,x_\sigma)$. Take, for example,
the star-unitary transformation
\begin{align}
  \label{LargeStar}
  u_{\mathbb{P},b}(x)=
  [1-\mathbb{P}(x_\rho,x_\sigma)]
  +e^{ib(x_\theta)}\mathbb{P}(x_\rho,x_\sigma),
\end{align}
where $b(x_\theta)$ is a real function. Its index $w(u)$ is
\begin{align}
  \begin{split}
    w(u_{\mathbb{P},b})
    ={}&
    \frac{i}{2\pi^2}
    \int \dd^3x\, b'(x_\theta)\cos^2\frac{b(x_\theta)}{2}
    (\mathbb{P}\star[\partial_\rho\mathbb{P},\partial_\sigma\mathbb{P}]_\star)
    \doteq
    \frac{1}{\pi}
    {\cal W}(\mathbb{P})
    \int \dd^3x\, b'(x_\theta)\cos^2\frac{b(x_\theta)}{2},
  \end{split}
\end{align}
the directions $(\rho,\sigma,\theta)$ have been defined in
footnote~\ref{foot:3dim}.
\textit{E.~g.} choosing $\mathbb{P}\doteq 2\exp \{-(x_\rho^2+x_\sigma^2)/|\theta|\}$ and
$b(x_\theta)|_{-\infty}^{+\infty}=2\pi$, we find $w(u)=1$. The
Lagrangian is not invariant under large star-gauge
transformations, but the path integral is, provided
\begin{align}
    \forall u\;
    \exists k\in\mathbb{Z}:\;
    \LL(\frac{m_g}{g^2}\RR)
    4\pi^2w(u)
    =2\pi k.
\end{align}
In the commutative theory, this enforces the quantization of the
dimensionless coupling constant $g^2/m_g$.

In three dimensions both Dirac and Majorana fermions exist.
Real fermions are specially interesting in that, for arbitrary values of the parameters
$\theta$, $g$, $m_f$, and $m_g$, our action~\eqref{symcs_action}
has a softly broken, ${\cal N}=1$ supersymmetry.
In the special case $m_f=m_g$ the action becomes invariant under the supersymmetry
transformations parameterized by $\epsilon$:
\begin{align}\label{susy_transformations}
    \delta A_\mu
    =
    i\bar\varepsilon \gamma_\mu \lambda
     ,
    \qquad
    \delta \lambda
    =
    \frac{1}{2} F_{\mu\nu}\gamma^\mu\gamma^\nu \varepsilon;
\end{align}
the supersymmetry algebra closes on star-gauge transformations. By
softly breaking supersymmetry we will be able to retrieve several
interesting limits, such as U$(1)_\star$ QCD ($m_g\rightarrow 0$,
$m_f$ finite), and pure Yang--Mills--Chern--Simons ($m_f\rightarrow\infty$, $m_g$
finite).

In the non-commutative setup, Lorentz symmetry is wasted. The
residual symmetry group can be obtained by considering the
Poincar\'e dual of $\theta_{\mu\nu}$,
\begin{align}
    \tilde \theta_{\lambda}
    \doteq
    \epsilon_{\lambda\mu\nu}\theta^{\mu\nu}
\end{align}
When $\tilde\theta_\lambda$ is timelike one has purely spacelike
non-commutativity, and the residual symmetry group is the
Galilei group of SO(2) rotations (space and time translations are
untouched by non-commutativity of the coordinates).
When $\tilde\theta_\mu$ is spacelike one has spacetime non-commutativity,
and the relic of the Lorentz group is SO(1,1). In the lightlike case
the entire Lorentz group is ruined.
In the following, we will consider the case where non-commutativity is spacelike.

Since Lorentz symmetry is broken one may try to construct new
tensor structures with the same spacetime, discrete and gauge
symmetry properties as the ones we have included in the Lagrangian
\eqref{symcs_action}. However, no \textit{perturbatively
renormalizable} such structures are induced in the effective
action by quantum effects. An interesting example is
\(
  \frac{1}{g}\int \dd^3x \; \epsilon^{\mu\nu\rho}\theta_{\mu\nu} A_\rho,
\)
which is gauge invariant, ${\cal P}$ and ${\cal T}$-odd and
${\cal PT}$-even, just like the Chern--Simons coefficient,
and would contribute a constant curvature
\(
  {\cal D}_\mu F^{\mu\nu}
  +\tilde F^\nu
  =
  \tilde \theta^\nu
\)
to the equations of motion,
destabilizing the perturbative vacuum already at tree level \cite{Sheikh-Jabbari:2001au}.
Such a term is however not generated by the Lagrangian
\eqref{symcs_action} at any order of perturbation theory, so we are
free to discard it.

Let us briefly recall the discrete symmetries of the theory.
In two space dimensions, a reflection with respect to a point can be
undone by a rotation, so the parity transformation ${\cal P}$ must be defined
as a reflection along \textit{one} space axis, say $x$
\begin{align}
  {\cal P}^{-1}A_\mu(t,x,y){\cal P}
  ={}&
  \begin{cases}
    -A_\mu(t,-x,y) & \text{if } \mu=1 ,
    \\
    \phantom{+}A_\mu(t,-x,y) & \text{otherwise} ;
  \end{cases}
\end{align}
charge conjugation
and time reversal act in the more familiar manner.
Direct inspection reveals that the NC Chern--Simons Lagrangian
is odd under parity and time-reversal.
While the fermion's kinetic term is even under the discrete transformations,
the fermion's mass term shares the same properties as the Chern--Simons mass term,
the two being related by supersymmetry~\eqref{susy_transformations}.
The whole Lagrangian is even under charge conjugation
and the combined action of ${\cal P}$ and ${\cal T}$;
the theory is therefore invariant under ${\cal CPT}$.

Since the mass terms share the same symmetry properties,
an anomalous Chern--Simons term can also be induced by massless fermions.
In this case, in \cite{Chu:2000bz}, a discontinuity was observed in the commutative $(\theta\rightarrow 0)$ limit,
related to the fact that Majorana spinors become non-interacting in this limit.
Since the Chern--Simons term stays finite for all values of $\theta$, however,
this discontinuity can be employed to test the presence of non-commutativity and
of violations of Lorentz invariance in nature.


\section{One-loop renormalization}\label{sec:radiative_corrections}

In the following we will address the issues of vacuum stability and
unitarity by analyzing the one-loop, one-particle irreducible two-point functions
of the gauge boson and Majorana fermion.
Our first task will be the calculation of the boson self-energy:
because  of the breaking of the Lorentz invariance,
the question of the transversality  of the self-energy is non-trivial, like in the finite-temperature case.
We will derive a Ward identity which shows the one-loop transversality.
We will be working in the Landau gauge, but we will discuss the gauge-independence of our result.
We find an unexpected IR/UV mixing, no unitarity violation for spacelike non-commutativity,
and an ``exotic'' low-dimensional remnant of the birefringence phenomenon.

\begin{figure}
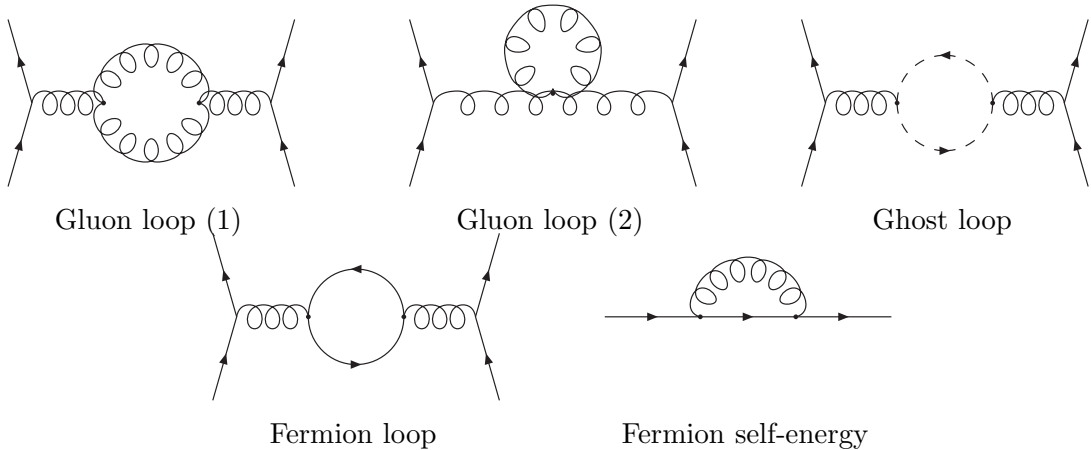

  \begin{center}
  \hspace{-1cm}
  \ScaledGlueLoop{0.9}{0}{0}{Gluon loop (1)}
  \ScaledTadpole{0.9}{0}{0}{Gluon loop (2)}
  \ScaledGhostLoop{0.9}{0}{0}{Ghost loop}
  \ScaledMajoranaLoop{0.9}{0}{0}{Fermion loop}
  \ScaledMajoranaSelfEnergy{0.9}{0}{0}{Fermion self-energy}
  \vspace{.5cm}
  \end{center}
  \caption{
    \linespread{0.9}\small\sf
    \label{fig:all_diagrams} All the relevant planar diagrams;
      all of these diagrams have non-planar counterparts,
      which contain the truly non-commutative effects.
      External particles have been added for future reference.
  }
\end{figure}

In a non-commutative theory, the momentum-dependent phases
in the vertices (see appendix~\ref{sec:FeynmanRulez})
factor out in the planar diagrams\footnote{
  The second diagram in figure~\ref{fig:all_diagrams},
  generated by a gluon four-vertex,
  is not present in $d=4$ theories, where it would contribute a non-gauge-invariant mass term.
  In our case, it plays an important dynamical role because of the breaking of Lorentz invariance.
} of figure~\ref{fig:all_diagrams}.
However, in the non-planar diagrams these phases retain a non-trivial dependence on the loop momentum,
and are a potential source of IR/UV mixing \cite{Minwalla:1999px}.

Let us recall some relevant features of the matterless model. The
commutative YMCS model is perturbatively finite, provided one
employs a gauge- and Lorentz-invariant regulator. By naive power
counting, at one loop there can be at most linear and logarithmic UV
divergences in the diagrams of figure~\ref{fig:all_diagrams}.
However, careful inspection of the vacuum polarization reveals that
the would-be linear divergence is generated by a parity-even
structure in the external momentum. If this divergence were
produced, it would contribute a parity-even, gauge-invariant,
power-counting renormalizable term to the effective action --- but
no Lorentz-invariant such term exists. Similarly, the theory is
protected against logarithmic divergences, as one easily sees by
checking that the coupling constant $g$ is dimensionful.

In the non-commutative case, non-planar diagrams have an even more
regular UV behavior, so one could expect perturbative finiteness
also here. This naive reasoning would lead one to not expect any
infrared/ultraviolet (IR/UV) mixing to occur: there is just no
divergence to mix with. In this section, however, we will see that
in the non-commutative case the loss of Lorentz-invariance  will
accomodate a linear IR/UV effect.

This is why supersymmetry turns out to be useful.
An elegant  way to have under control the IR/UV effects is to
consider the supersymmetric  extension of the model:
SUSY provides a natural, gauge-invariant regularization and
acts via the IR/UV mixing as an infrared regulator~\cite{Alvarez-Gaume:2003mb}.
In fact, if the number of supersymmetries is sufficiently large, all
the undesired divergences will disappear from the infrared region.
The sum of the first four diagrams in figure~\ref{fig:all_diagrams} is related by the supersymmetry
Ward--Takahashi identities to the single fermion self-energy diagram.
As some trivial power counting shows, the polarization tensor potentially
has linear divergences, but the fermion's self-energy has at most
a logarithmic divergence, which shows that (unless
non-commutativity breaks SUSY) the fermion loop diagram must
cancel the linear divergences of the purely gauge/ghost diagrams.

\subsection{Tensor structures  on $\mathbb{R}^3_\theta$}

The simplest way to address the question of vacuum stability and unitarity
is to analyze the one-loop, one-particle irreducible two-point function of the gauge boson.
At the tree level, this function coincides with the commutative one, since
the star-product is irrelevant in the quadratic part of the action.
Its tree-level form in the Landau gauge (which we shall always use in the following) is in fact
\begin{equation}
  \label{Gammatree}
  \Gamma^{\mathrm{tree}}_{\mu\nu}(p)
  =
  {{{g}}}_{\mu\nu} p^2-p_\mu p_\nu-i m_g \epsilon_{\mu\nu\lambda} p^\lambda.
\end{equation}
In the commutative case
the only effect of  the one-loop radiative corrections is to
renormalize the two transverse structures in (\ref{Gammatree}):
\begin{equation}
\label{Gammaoneloop}
  \Pi_{\mu\nu}(p)
  =
  \Pi^{\rm e}(p)({{{g}}}_{\mu\nu} p^2-p_\mu p_\nu)
  -i m_g \Pi^{\rm o}(p)\epsilon_{\mu\nu\lambda} p^\lambda.
\end{equation}
This simple setting cannot be promoted to the
non-commutative setup as it originates from the simultaneous
presence of gauge- and Poincar\'e-invariance, the last of which is now broken.
More importantly, even the transversality of the one-loop
correction to the $\Gamma^{\mathrm{tree}}_{\mu\nu}$ may be endangered.
This possibility, for example, is realized in non-Abelian gauge theories at finite temperature
\cite{Gross:1980br,Weldon:1996kb}, where the spacetime symmetries are
destroyed by the existence of a preferred reference frame, provided by the thermal bath.

Because of the presence of the noncommutativity parameter
$\theta_{\mu\nu}$, it is possible to construct more tensor
structures than in the commutative case. By introducing
$\tilde\theta^\mu=\frac{1}{2}\epsilon^{\mu\rho\sigma}\theta_{\rho\sigma}$
and the
orthogonal basis $(p^\mu,\tilde p^\mu,\chi^\mu)$, with
\begin{eqnarray}\label{OrthogonalBasis}
  \chi^\mu=\tilde\theta^\mu -\frac{(\tilde \theta\cdot p)}{p^2} p^\mu
  ,
  \qquad
  \tilde p^\mu
  =\epsilon^{\mu\rho\sigma}\chi_\rho p_\sigma
  =\epsilon^{\mu\rho\sigma}\tilde\theta_\rho p_\sigma
  =\theta^{\mu\sigma}p_\sigma
  =(\theta p)^\mu
  ,
\end{eqnarray}
which satisfies the  completeness relation
\begin{equation}
  \frac{\chi_\mu\chi_\nu}{(\chi\cdot\chi)}+ \frac{\tilde p_\mu
  \tilde p_\nu}{(\tilde p\cdot \tilde p)}+\frac{p_\mu
  p_\nu}{(p\cdot p)}={{{g}}}_{\mu\nu}
  ,
\end{equation}
we can write the most general expression for the self-energy
\begin{equation}
\Pi_{\mu\nu}= (A_1 \chi_{\mu}+A_2 \tilde p_\mu+A_3 p_\mu)\chi_\nu+
(B_1 \chi_{\mu}+B_2 \tilde p_\mu+B_3 p_\mu)\tilde p_\nu+(C_1
\chi_{\mu}+C_2 \tilde p_\mu+C_3 p_\mu)p_\nu.
\end{equation}
It is possible to reduce the number of allowed tensor structures using the appropriate Ward identity for the gluon propagator.
We start by considering the  BRST variation\footnote{
  The fields transform as follows:
  \[
    \delta A_\mu = {\cal D}^\star_\mu c \,\epsilon
    ,
    \qquad
    \delta c = -ig (c\star c) \,\epsilon
    ,
    \qquad
    \delta \bar c = \frac{1}{\xi} \partial_\mu A^\mu \,\epsilon
    ,
  \]
  where $c$ is the ghost field, $\epsilon$ is the Grassman BRST parameter, and $\xi$ is the gauge-fixing parameter.
}
of the correlator  $\langle \bar c(x)
A_\nu (y)\rangle_0$:
\begin{eqnarray}
  \label{BRST1}
  0
  =
  \delta(\langle \bar c(x) A_\nu (y)\rangle_0)
  =
  \frac{1}{\xi}\langle\partial_\mu A^{\mu}(x)
  A_\nu(y)\rangle_0+\langle \bar c(x)\partial_\nu c(y)\rangle+g
  \langle \bar c(x)[A_\nu(y),c(y)]_\star\rangle_0
  ;
\end{eqnarray}
by employing   also the Schwinger--Dyson  equation for the ghost propagator
\begin{equation}
  \square_x {\cal G}(x-y)+g \langle
  \partial_{\nu} [A^\nu(x),c(x)]_\star \bar c(y)\rangle_0=
    \delta^3(x-y),
\end{equation}
we  obtain the identity
\begin{eqnarray}
  p_\lambda \, i\Pi^{\lambda\alpha}(p)
  =
  ig   \Gamma_\nu(p)\left(p^\nu p^\alpha-p^2\delta^{\nu\alpha}
  +im_g\epsilon^{\nu\alpha\beta} p_\beta
  -i\Pi^{\nu\alpha}(p) \right)
  ,
\end{eqnarray}
where   $ \Gamma_\nu $ is defined by:
\begin{equation}\label{define_may_gammas}
 \langle \bar c(x) [A_\nu(y),
 c(y)]_\star \rangle_0\equiv i \int d^3 z \mathcal{G}(x-y-z) \Gamma_\nu (z)\equiv i \tilde \Gamma_\nu(x-y),
\end{equation}
and $ {\cal G}(x-y)\doteq \langle c(x) \bar c(y)\rangle$ is the ghost
propagator.
At one loop, the gluon self-energy $\Pi^{\mu\nu}$ is of order
$g^2$, while $\Gamma_\nu(p)$ is of order $g$. Then the BRST
identity reduces to
\begin{equation}\label{BRST4}
  p_\lambda\,i\Pi^{\lambda\nu}_{1-loop}
  =
  ig\Gamma^{g}_\nu(p)\left(p^\nu
  p^\alpha-p^2\delta^{\nu\alpha}
  +im_g \epsilon^{\nu\alpha\beta}p_\beta\right)
  .
\end{equation}
To fully appreciate the meaning of this constraint, we have to
analyze the form of $\Gamma^{g}_\nu(p)$, the function
$\Gamma_\nu(p)$ at order $g$ in the coupling constant: at this
order we may use the bare ghost propagator inside
equation~\eqref{define_may_gammas}:
\begin{equation}
  \frac{i}{p^2}\Gamma^{g}_\nu (p)= \tilde\Gamma_\nu(p).
\end{equation}

In the commutative case, $\Gamma_\nu $ is compelled by Lorentz invariance  to be
proportional to $p_\nu$, and the above identity entails
transversality.
As we have seen in equation~\eqref{OrthogonalBasis},
in the non-commutative model, there are two new
possible vectors that can appear in the expansion of
$\Gamma_\nu$,
and the above argument would seem to fail. Nevertheless, a detailed one-loop
analysis shows that $\Gamma^\nu$ has surprisingly no component
along $\tilde p^\mu$ and $\chi^\mu$, so transversality is preserved at one loop.
At higher loops the situation is less clear, but we have indications that this property
continues to hold.

Imposing $p^\mu\Pi_{\mu\nu}=0$ and
subsequently $p^\nu\Pi_{\mu\nu}=0$, we immediately get that all the $C_i$ vanish, and that $A_3=B_3=0$.
Therefore, the most general transverse two-tensor is
\begin{eqnarray}
  \label{self}
  \Pi_{\mu\nu}
  &={}&\nonumber
  (A_1 \chi_{\mu}+A_2 \tilde p_\mu)\chi_\nu
  + (B_1\chi_{\mu}+B_2 \tilde p_\mu)\tilde p_\nu
  =\\
  &\doteq&\nonumber
  \Pi^{\rm e}_1 p^2
  \left ({{{g}}}_{\mu\nu}-\frac{p_\mu p_\nu}{p^2}-\frac{\tilde p_\mu\tilde p_\nu}{\tilde p^2}\right)
  + \Pi^{\rm e}_2 \frac{\tilde p_\mu\tilde p_\nu}{\tilde p^2} p^2
  -\Pi^{\rm o} im \epsilon_{\mu\nu\lambda}p^\lambda
  +\bar\Pi^{\rm o}(\tilde p_\mu \chi_\nu+\tilde p_\nu \chi_\mu)
  .
\end{eqnarray}
In practice the last tensor structure will not be generated at any
order in perturbation theory, because of the accidental invariance
$\theta\to -\theta$ possessed by $\Pi_{\mu\nu}$; this, combined with
the Bose--Einstein symmetry ($\mu \rightarrow \nu,p \rightarrow
-p$), would require $\bar\Pi^{\rm o}$ to be even in $\theta $ and
odd in $p$, however, no such scalar can be built using only $p^2$,
$\tilde\theta^2$ and $\tilde\theta p$. We are left with
\begin{align}
\begin{split} \label{self-1}
  \Pi_{\mu\nu}={}&
  \Pi^{\rm e}_1 \frac{\chi_\mu \chi_\nu}{\chi^2} p^2
  + \Pi^{\rm e}_2 \frac{\tilde p_\mu\tilde p_\nu}{\tilde p^2} p^2
  -\Pi^{\rm o} im_g \epsilon_{\mu\nu\lambda}p^\lambda
  =\\={}&
  \Pi^{\rm e}_1 \left({{{g}}}_{\mu\nu}-\frac{p_\mu p_\nu}{p^2}-\frac{\tilde p_\mu\tilde p_\nu}{\tilde p^2}\right) p^2
  + \Pi^{\rm e}_2 \frac{\tilde p_\mu\tilde p_\nu}{\tilde p^2} p^2
  -\Pi^{\rm o} im_g \epsilon_{\mu\nu\lambda}p^\lambda
  =\\={}&
  \Pi^{\rm e}_1 \left({{{g}}}_{\mu\nu}-\frac{p_\mu p_\nu}{p^2}\right) p^2
  + \Pi^{\theta} \frac{\tilde p_\mu\tilde p_\nu}{\tilde p^2} p^2
  -\Pi^{\rm o} im_g \epsilon_{\mu\nu\lambda}p^\lambda;
\end{split}
\end{align}
in the last line, we have explicitly isolated the contribution from the
new tensor structure $\Pi^{\theta}\doteq\Pi^{\rm e}_2-\Pi^{\rm e}_1$.
At the end of the day, the only effect of non-commutativity is to produce two different
wavefunction renormalizations: one along $\tilde p_\mu$
and one along $\chi_\mu$.
Because of
\[
  {{{g}}}_{\mu\nu}-\frac{p_\mu p_\nu}{p^2}=
 \frac{\chi_\mu \chi_\nu}{\chi^2}+\frac{\tilde p_\mu\tilde p_\nu}{\tilde p^2},
\]
the commutative case (\ref{Gammaoneloop}) is recovered when $\Pi^{\rm e}_1=\Pi^{\rm e}_2$.

\subsection{One-loop polarization tensor}\label{sec:polarization_tensor}

In the next section, by inspecting the renormalized propagator
$G^R_{\mu\nu}(p)$ at one loop, we shall illustrate how
non-commutativity affects the spectrum of the theory, its
unitarity and vacuum stability.
But for accomplishing that, we need the explicit form of scalar functions $\Pi^{\rm e}_1$,
$\Pi^{\rm e}_2$ and $\Pi^{\rm o}$, whose evaluation is quite lengthy and tedious.
The techniques employed are well known in the literature, see for example \cite{Brandt:2001ud}.
Each renormalization function contains two contributions, which originate
from the planar and non-planar diagrams.
The former is identical\footnote{Up to overall
   factors, related for example to the fermions being
  Majorana and not Dirac.}
to the commutative (non-Abelian) one, while the latter
carries the effects of non-commutativity.

It is usually convenient to measure everything units of $m_g$, so that the
action's parameters $(m_f,m_g,g,\theta)$ are re-expressed in terms of the dimensionless quantities
$\LL(\mu,g^2/m_g,m_g^2\theta\RR)$.
\begin{align}
  \begin{split}
     \label{DimensionlessUnits}
     \mu \doteq \frac{m_f}{m_g}
     ,
     \qquad
     \eta \doteq{}\frac{\sqrt{p^\mu p_\mu}}{m_g}
     ,
     \qquad
     \xi  \doteq{} m_g\sqrt{p\bullet p} ={} m_g \theta |\vec p| = m_g |\tilde p|
     .
   \end{split}
\end{align}
We start by evaluating the contribution to the gluon self-energy
from the gauge sector's contributions. We calculate the
diagrams and project according to the convention~\eqref{self}. The
planar contributions are
\begin{eqnarray*}
  i\Pi^{\rm e,glue}_{\rm 1,pl}
  =
  i\Pi^{\rm e,glue}_{\rm 2,pl}
  &=&
  \frac{i g^2}{32\pi m_g}
  \Bigg\{
  5-\frac{11}{\eta^2}-\frac{|\eta|}{2\eta^4}
  \Bigg[
    (\eta^4+13\eta^2+4)(4-\eta^2)\sin^{-1}\left(\frac{\eta^2}{\eta^2-4}\right)^{\frac{1}{2}}
    +\\&&
    +(\eta^2+7)(\eta^4+1)^2\sin^{-1}\left(\frac{\eta^2+1}{\eta^2-1}\right)
    +\pi\left(2\eta^4-\frac{13}{2}\eta^2+\frac{7}{2}\right)
  \Bigg]
  \Bigg\}
  ,
\\
  i\Pi^{\rm o,glue}_{\rm pl}
  &=&
  -\frac{ig^2}{16\pi m}
  \Bigg\{2-\frac{1}{\eta^2}-\frac{|\eta|}{4\eta^4}
  \Bigg[
  -6\eta^2(\eta^2+2)(4-\eta^2)  \sin^{-1}\left(\frac{\eta^2}{\eta^2-4}\right)^{\frac{1}{2}}
  +\\&&
  -2(3\eta^2+1)(1-\eta^2)^2\sin^{-1}\left(\frac{\eta^2+1}{\eta^2-1}\right)
  +\pi(\eta^4-\eta^2-1)
  \Bigg]
  \Bigg\}
  .
\end{eqnarray*}
Clearly, $\Pi^\theta_{\rm pl}=\Pi^{\rm e}_{\rm 2,pl}-\Pi^{\rm e}_{\rm 1,pl}$ vanishes.
The non-planar contributions are
\begin{eqnarray*}
  i\Pi^{\rm e,glue}_{\rm 1,np}
  &=&
  \frac{ig^2}{8\pi m_g}\int_0^1 \dd x\Bigg[
  \LL(
    \frac{(4-5\eta^2+\eta^4)}{\eta^2 \xi}
    -\frac{9(4-\eta^2)}{4\sqrt{x(x-1)\eta^2+1}}
  \RR)e^{-\xi\sqrt{x(x-1)\eta^2+1}}
  +\\&&
  +\LL(
     \frac{(6-2\eta^2)}{\xi}
     -\frac{ 5(1-\eta^2)^2}{2\eta^2\sqrt{x(x-1)\eta^2+x}}
  \RR)e^{-\xi\sqrt{x(x-1)\eta^2+x}}
  +\\&&
  +\LL(
    \frac{\eta^2}{4\sqrt{x(x-1)\eta^2}}
    -\frac{(1-\eta^2)}{\xi}
  \RR)e^{-\xi\sqrt{x(x-1)\eta^2}}
  +\left(\frac{(5-4\eta^2)-(9-4\eta^2)e^{-\xi}}{\eta^2\xi}\right)
  \Bigg]
  ,
\\
  i\Pi^{\rm e,glue}_{2,\rm np}
  &=&
  \frac{ig^2}{8\pi m_g}\int_0^1 \dd x
  \Bigg[-\Bigg(\frac{9(4-\eta^2)}{4\sqrt{ x(x-1)\eta^2+1}}
  +\frac{(4-5\eta^2+\eta^4)}{\eta^2}\sqrt{x(x-1)\eta^2+1} \Bigg)
  e^{-\xi\sqrt{x(x-1)\eta^2+1}}
  +\\&&
  -\Bigg(  \frac{5(1-\eta^2)^2}{2\eta^2\sqrt{x(x-1)\eta^2+x}}
  + (6-2\eta^2)\sqrt{x(x-1)\eta^2+x}  \Bigg)   e^{-\xi\sqrt{x(x-1)\eta^2+x}}
  \\&&
  -\Bigg(-(1-\eta^2)\sqrt{x(x-1)\eta^2} -\frac{\eta^2}{4\sqrt{x(x-1)\eta^2}}   \Bigg)
  e^{-\xi\sqrt{x(x-1)\eta^2}}
  +\Bigg(\frac{(5-4\eta^2)-(9-4\eta^2)e^{-\xi} }{\eta^2\xi}\Bigg)\Bigg]
  ,
\\
  i\Pi^{\rm  o,glue}_{\rm np}
  &=&
  -\frac{ig^2}{16\pi m}\int_0^1 \dd x
  \Bigg[
  (1-\eta^2)(2x-3(1-\eta^2))\frac{e^{-\xi\sqrt{x(x-1)\eta^2+x}}}{\sqrt{x(x-1)\eta^2+x}}
  -\frac{e^{-\xi\sqrt{x}}}{\sqrt{x}}
  +\\&&
  +\Bigg(\frac{3}{2}\eta^4-2\eta^2\Bigg)\frac{e^{-\xi\sqrt{x(x-1)\eta^2}}}{\sqrt{x(x-1)\eta^2}}
  +\Bigg(\frac{3}{2}\eta^4-3\eta^2-12\Bigg)\frac{e^{-\xi\sqrt{x(x-1)\eta^2+1}}}{\sqrt{x(x-1)\eta^2+1}}
  \Bigg]
  .
\end{eqnarray*}
A few remarks are required  at this point. First of all, we observe
that the polarization tensor is a combination  of those transverse
tensor structures that we expected from the Ward identity.
Second, since the diagrams contributing to the self-energy are the
same that one has to compute in commutative topologically-massive
QCD, we know that the planar contribution has to be equal to the one
in~\cite{Pisarski:1985yj}. We display explicitly the planar
contribution, so one can immediately verify that this is indeed the
case.
Finally, since star-commutators vanish in the commutative limit, one
expects the planar and non-planar contributions to cancel against
each other as $\theta\rightarrow 0$. We observe that $\Pi^{\rm
o,glue}_{\rm np}$ and $\Pi^{\rm e,glue}_{\rm 1,np}$ are regular in
the limit $\theta \rightarrow 0$ and exactly cancel their planar
counterparts in this limit. $\Pi^{\rm e,glue}_{\rm 2,np}$ is instead
singular in the $\theta \rightarrow 0$ as well as in  the $|\vec
p|\rightarrow 0$ limits. This is the characteristic signature of an
IR/UV mixing of divergences.

As we remarked in the introduction to this section, this IR/UV
mixing is somewhat unexpected, given that the commutative theory is
pertrubatively finite. However, one can see that the diagrams
contributing to the self-energy are power-counting divergent and
that, in the commutative case, they produce a finite result only when
evaluated with a gauge-invariant and Lorentz-invariant regulator.

We can thus understand the birth of the IR divergences in the
non-commutative case by considering that the  effective cut-off
$\Lambda_{\rm eff}^{-1}=\theta |\vec p|$ simulates the effect of  a
non-Lorentz-invariant regulator, which, when removed ($\Lambda_{\rm
eff}\rightarrow\infty$), reintroduces the original divergences in
the infrared ($\theta |\vec p|\rightarrow 0$) domain.
All the new infrared divergences are contained in the Lorentz-breaking tensorial structure
multiplying $\Pi_\theta=(\Pi^{\rm e}_2-\Pi^{\rm e}_1)$ in equation \eqref{self-1}:
the symmetries of the commutative theory prevent such a term from being generated.

These are important checks of the consistency of our calculations.
However, the fact that we are working in the Landau gauge
might lead one to suspect this IR/UV phenomenon to be a gauge artifact.
This problem was investigated in \cite{Ruiz:2000hu};  in section~\ref{sec:why_they_occur}
we will make some further remarks on these issues.

For what concerns the contribution from the fermion loop
we proceed as we did before, and we find that the
fermion contribution indeed has the structure (\ref{self}).
The planar part is
\begin{align}
  \begin{split}
  \label{majo_pol_1}
    i\Pi^{\rm e,ferm}_{\rm 1,pl}
    =
    i\Pi^{\rm e,ferm}_{\rm 2,pl}
    ={}&
    \phantom{+}
    \frac{ig^2}{64\pi m_g}
    \LL[
      \frac{4{\mu}}{\eta^2}
      +\frac{(\eta^2+4\mu^2)}{\eta^3}\log\LL(\frac{2{\mu}-\eta}{2{\mu}+\eta}\RR)
    \RR],
    \\
    i\Pi^{\rm o,ferm}_{\rm pl}
   ={}&
   -
   \frac{ig^2}{4\pi m_g}
    \frac{\mu}{\eta}\log\LL(\frac{2{\mu}-\eta}{2{\mu}+\eta}\RR)
    .
  \end{split}
\end{align}
The non-planar part is
\begin{align}
  \begin{split}
  \label{majo_pol_2}
    i\Pi^{\rm e,ferm}_{\rm 1,np}
    ={}&\phantom{-}
    \frac{ig^2}{2\pi m_g}
    \int_0^1 \dd x\,
    \frac{x(1-x)}{\sqrt{\mu^2+x(x-1)\eta^2}}
    e^{-\xi \sqrt{\mu^2+x(x-1)\eta^2}}\,
    ,
\\
    i\Pi^{\rm e,ferm}_{\rm 2,np}
    ={}&\phantom{-}
    \frac{ig^2}{2\pi m_g}
    \int_0^1 \dd x
    \Bigg[
    \frac{1}{\eta^2}\Bigg(\sqrt{\mu^2+x(x-1)\eta^2}+
    \frac{1}{\xi}\Bigg)e^{-\xi \sqrt{\mu^2+x(x-1)\eta^2}}
\\ &
    \phantom{-\frac{ig^2}{2\pi m_g}\int_0^1 \dd x\Bigg[}
    +\frac{x(1-x)}{\sqrt{\mu^2+x(x-1)\eta^2}}e^{-\xi \sqrt{\mu^2+x(x-1)\eta^2}}
    \Bigg],
\\
    i\Pi^{\rm o,ferm}_{\rm np}
    ={}&
    -    \frac{ig^2}{4\pi m_g}
    \int_0^1 \dd x\,
    \frac{\mu}{\sqrt{\mu^2+x(x-1)\eta^2}}\,
    e^{-\xi\sqrt{\mu^2+x(x-1)\eta^2}}
    .
  \end{split}
\end{align}
The fermionic contributions display a behaviour that is analogous to
the gauge ones. However, the infrared divergences occur here with an
opposite sign: hence, the full contributions to the polarization
tensor
\begin{align}
\begin{split}
  \Pi^{\rm e}_1 ={}& \LL( \Pi^{\rm e,glue}_{\rm 1,np} + \Pi^{\rm e,glue}_{\rm 1,pl} \RR) + \LL( \Pi^{\rm e,ferm}_{\rm 1,np} + \Pi^{\rm e,ferm}_{\rm 1,pl} \RR)
  ,
  \\
  \Pi^{\rm e}_2 ={}& \LL( \Pi^{\rm e,glue}_{\rm 2,np} + \Pi^{\rm e,glue}_{\rm 2,pl} \RR) + \LL( \Pi^{\rm e,ferm}_{\rm 2,np} + \Pi^{\rm e,ferm}_{\rm 2,pl} \RR)
  ,
  \\
  \Pi^{\rm o} ={}& \LL(\Pi^{\rm o,glue}_{\rm np} + \Pi^{\rm o,glue}_{\rm pl}\RR) + \LL(\Pi^{\rm o,ferm}_{\rm np} + \Pi^{\rm o,ferm}_{\rm pl}\RR)
\end{split}
\end{align}
are all finite and vanish as $\theta |\vec p|\rightarrow 0$. We see
clearly that supersymmetry acts as a gauge-invariant regulator
smoothing out the IR divergences.

\subsubsection{Renormalization and analytical structure}

In the commutative case, the two functions $\Pi^{\rm e}$ and $\Pi^{\rm o}$
of equation~\eqref{Gammaoneloop} govern the wavefunction and mass renormalization respectively
\cite{Deser:1981wh,Deser:1982vy,Pisarski:1985yj}.
Once Lorentz invariance is lost, however, we cannot  expect just one
wavefunction ($\mathcal{Z}_{e}=1-\Pi^{\rm e}$) and mass
($\mathcal{Z}_{m}=1-\Pi^{\rm o}$) renormalization, since as we have seen the different
components of the gauge field renormalize in different ways.

In the non-commutative case, we have to arrange the contributions to the self energy
introducing the renormalization functions and renormalized mass (defined as the position
of the physical pole in the propagator) as
\begin{align}
  \begin{split}
  \mathcal{Z}_1 ={}& (1-\Pi_1^{\rm e})
  ,
  \\
  \mathcal{Z}_2 ={}& (1-\Pi_2^{\rm e})
  ,
  \\
  \mathcal{Z}_m ={}& (1-\Pi^{\rm o})
  ,
  \\
  (m^{\rm R}_g)^2 ={}& \LL(\frac{\mathcal{Z}_m^2}{\mathcal{Z}_1\mathcal{Z}_2}\RR)m_g^2
  ;
  \end{split}
\end{align}
and we get the renormalized propagator:
\begin{eqnarray}\nonumber
  G^R_{\mu\nu}(p)
  =
  \frac{i}{\sqrt{\mathcal{Z}_1 \mathcal{Z}_2}(p^2-(m^{\rm R}_g)^2) }
  \Bigg(
  \sqrt{\frac{\mathcal{Z}_2}{\mathcal{Z}_1}}
  \left({{{g}}}_{\mu\nu}-\frac{p_\mu p_\nu}{p^2}
  -\frac{\tilde p_\mu\tilde p_\nu}{\tilde p^2}\right)
  +\sqrt{\frac{\mathcal{Z}_1}{\mathcal{Z}_2}}\frac{\tilde p_\mu\tilde p_\nu}{\tilde p^2}
  + im^{\rm R}_g\epsilon_{\mu\nu\lambda}\frac{p^\lambda}{p^2} \Bigg)
  .
\end{eqnarray}
The propagator contains an unphysical pole as a result of the propagation of the massless degree of freedom,
as well as a physical pole for $p^2=(m_g^{\rm R})^2.$
To order $g^2/m_g$ we have:
\begin{eqnarray}
 \nonumber (m_g^{\rm R})^2
  =
  m_g^2\frac{(1-\Pi^{\rm o})^2}{(1-\Pi^{\rm e}_1)(1-\Pi^{\rm e}_2)}
  =
  m_g^2[1-2\Pi^{\rm o}+\Pi^{\rm e}_1+\Pi^{\rm e}_2 ]
  +{\rm O}(g^2/m_g^4)
  .
\end{eqnarray}
We need to prove that the renormalized mass does not depend on the gauge.

To show this, we can proceed as follows\footnote{See
\cite{Pisarski:1985yj} for a similar analysis in the commutative
case}. First of all, we construct the polarization vector
$\epsilon^\nu$. In Landau gauge, the linearized Euler--Lagrange
equation for $A_\mu$ is:
\begin{equation}
   \square A^\nu +m_g \epsilon^{\nu\alpha\beta }\partial_\alpha A_\beta
   =
   0
   .
\end{equation}
Let us expand $A^\nu$ in plane waves as  $A^\nu=\epsilon^\nu
e^{ipx}$,  on shell\footnote{Just
  like in the commutative case, the
  Euler--Lagrange equations also have the massless solution
  $\epsilon^\mu\propto p^\mu$, but this can be gauged away.
}
($p^2=m_g^2$)  we find the equation
\begin{eqnarray}
  \epsilon^\nu
  =
  \frac{i}{m_g}\epsilon^{\nu\alpha\beta}p_\alpha
  \epsilon_\beta,
\label{polarizzo}
\end{eqnarray}
we can satisfy this equation in Landau gauge
($p^\mu \epsilon_\mu=0$) by choosing:
\begin{equation}
  \epsilon^\mu
  =
  \frac{\widetilde{p}_\mu}{\sqrt{m^2 \tilde\theta^2-(\tilde \theta \cdot p)^2}}-
  im_g \frac{\chi^\mu}{\sqrt{m_g^2 \tilde \theta^2 -(\tilde \theta \cdot p)^2}}
  \label{indi}
  .
\end{equation}
It is possible to construct a gauge-invariant observable by contracting the self energy
with the polarization tensors:
\begin{equation}
  \epsilon^{\dag\mu}(p) \Pi_{\mu\nu}\epsilon^\nu(p)
  =
  m^2 \Bigg(\Pi^{\rm e}_1+\Pi^{\rm e}_2   -2\Pi^{\rm o} \Bigg)_{p^2=m^2}
  =
  \Bigg((m^{\rm R}_g)^2-m_g^2\Bigg)
  \doteq
  \Delta m_g^2
  .
\label{fisica}
\end{equation}
This expression, evaluated on shell,  is  exactly the mass shift.
Our result is thus a physically observable quantity, and as such
it ought to be gauge independent.

By observing the mass-shift expression off-shell, we find
non-vanishing imaginary contributions beginning at the thresholds
$p^2=0$, $p^2=m_g^2$, and $p^2=4m_g^2$. The first cut indicates
the production of unphysical (zero mass) degrees of freedom, the
second one is related to the production of pure gauge modes
together with one physical gauge boson, and the third one to the
generation of two physical degrees of freedom (plus eventual massless modes).
One finds this analytical structure also in the commutative case.
On-shell, where all the unphysical gauge effects disappear,  we
observe that all the imaginary contributions cancel. We then
conclude that non-commutativity does not alter the analytical
structure of the theory; in particular, we observe no violation of
unitarity\footnote{
  Our considerations refer to the case of spacelike non-commutativity.
  Timelike non-commutativity seems to be inconsistent, as can be checked within our setup
  by considering the quantity $p\bullet p$, of equation~\eqref{pBulletp}:
  for timelike non-commutativity, one can have $p \bullet p<0$ for $p$ spacelike,
  and the amplitude develops an imaginary part irrespectively of the total value of $p^2$,
  signalling a violation of unitarity. See \cite{Gomis:2000zz,Bassetto:2001vf,Bahns:2002vm,Bahns:2003vb,Liao:2002pj}.
}.
In fact, the gauge-invariant  amplitude evaluated
on-shell (below the threshold for the production of two gauge
bosons) is real. Of course the amplitude develops an imaginary
part, due to the fermion loop, when $m_g=2 m_f$, associated to the
decay process $g\rightarrow ff$.

\subsubsection{Birefringence and renormalization of the Chern--Simons coefficient}

In order to write the one-loop expression of the propagator,
we had to introduce two renormalization functions $ \mathcal{Z}_1,
\mathcal{Z}_2$. This fact  is characteristic of spaces which
exhibit a breaking of Lorentz invariance.
In particular, in non-commutative QED this fact leads to a birefringence phenomenon:
the two physical degrees of freedom renormalize independently \cite{Guralnik:2001ax}.
In $d=2+1$, instead, there is only one physical polarization and we find that the
renormalized polarization vector which solves the one-loop corrected
equations of motion
\begin{equation}
  \left(
  {\cal Z}_1 \LL(
     {{{g}}}^{\mu\nu} p^2
     -p^\mu p^\nu
    -\frac{\tilde p_\mu\tilde p_\nu}{\tilde p^2} p^2
  \RR)
  +{\cal Z}_2\frac{\tilde p_\mu\tilde p_\nu}{\tilde p^2} p^2
  -im_g {\cal Z}_m
  \epsilon^{\mu\nu\lambda}p_\lambda
  \right)
  \epsilon^{\rm r}_\nu=0
\end{equation}
is rotated with respect to the bare one, which solves the tree-level e.o.m.
\begin{equation}
  \left({{{g}}}^{\mu\nu} p^2-p^\mu p^\nu - im_g \epsilon^{\mu\nu\lambda}p_\lambda\right)\epsilon_\nu
  =
  0.
\end{equation}
We stress that this happens not only because
of the breaking of Lorentz invariance, which implies $ \mathcal{Z}_1\neq\mathcal{Z}_2$,
but also owing to the presence the Chern--Simons term which couples
the polarizations $\widetilde{p}_\mu, \chi^\mu$.

As we have seen, in the commutative case the Chern--Simons action
is \textit{not} invariant under gauge transformations but, being a
topological invariant, it is still compatible with an invariant
partition function, provided that its variation is a multiple of
$2\pi$ --- this fixes the ratio between the coupling and mass to
be an integer.
The one-loop gauge-invariance of the partition function in the
commutative case is ensured if the following condition is satisfied:
\begin{align*}
 4\, \pi\, \left(
  \frac{m_g}{g^2}
  \right)_{\rm ren}
  =
  4 \, \pi \,Z_m \, \left(\frac{Z}{Z_g}\right)^2
  \left(
  \frac{m_g}{g^2}
  \right)_{\rm bare}=q
  ,
\end{align*}
where $q$ is a (positive) integer. In \cite{Pisarski:1985yj}, it has
been shown  that this condition is satisfied thanks to a topological
Ward identity that relates $Z_m$ to the other renormalization
constants of the theory. Subsequently, it
has been shown that the ratio ${m_g}/{g^2}$ is
unrenormalized  beyond one loop in all infrared-safe gauges \cite{Brandt:2000qp}.

In our case, the removal of the cut-off ($\mu\rightarrow\infty$)
will leave us with diverging renormalization constants (for {$|\vec
p|\rightarrow 0$}) and we  find $({m_g}/{g^2})_{\rm ren}\rightarrow
0$.
The more reasonable interpretation is that it is meaningless to
speak about the renormalization of the Chern--Simons coefficient,
since the theory is not IR-finite: the poles in the infrared forbid
the expansion of the effective action in  the momenta and the
identification of the linear term in $p$ as the one responsible for the
CS-renormalization.
This is to be contrasted with the pure CS theory, that is infrared-finite \cite{Bichl:2000bq}.
In that case, the renormalization of the CS coefficient is well defined
\cite{Sheikh-Jabbari:2001au,Chen:2000ak,Bak:2001ze,Martin:2001ci,Das:2001kf}.

\subsection{Majorana self-energy}\label{sec:majo_self}

The tensor structures appearing in the Majorana self-energy are
the ones encountered in the commutative case, and no additional
terms are induced\footnote{For example,
  the structure $\gamma^\mu\theta_{\mu\nu}k^\nu$ cannot be generated because it is odd in $\theta$.
}.
Therefore, we use the conventional parameterization:
\begin{equation}
  -i\Sigma
  =
  i\Sigma_\psi  \frac{\ssh k}{m_f}
  -i\Sigma_{m_f}\Id
\end{equation}
The computation again employs the standard techniques. The planar contributions
are
\begin{eqnarray*}
  \frac{\Sigma_{\psi,\rm p}}{m_g}
  &={}&
  \frac{g^2}{2\pi m_g}
  \frac{1}{4\eta^3}
  \Bigg[
  ( \eta^2-\mu^2 )
  [ \eta^2-\mu(2+\mu)]
  \log\LL( \frac{\mu^2-\eta^2}{(\eta+{\mu})^2} \RR)
  +\\&&\phantom{\frac{1}{4\eta^3}\Big\{}
  +  (\eta^2-\mu^2+1 )
  [ \eta^2-(1+\mu)^2 ]
  \log\LL( \frac{\mu^2-(\eta-1)^2}{(\eta+{\mu})^2-1} \RR)
  \Bigg]
  +\\&&
  +
  \frac{g^2}{2\pi m_g}
  \LL(\frac
  {  2{\mu}\LL[ \eta^2+\mu(2+\mu)+\fract{1}{2} \RR]
  +\LL[\eta^2-(1+\mu)^2\RR]  }
  {2\eta^2}\RR)
\\
  \frac{\Sigma_{m, \rm p}}{m_g}
  &={}&
  \frac{g^2}{2\pi m_g}
  \frac{1}{2\eta}
  \Bigg[
  (\eta^2-\mu^2) \log\LL( \frac{\mu^2-\eta^2}{(\eta+{\mu})^2} \RR)
  -[ \eta^2-(1+\mu)^2 ] \log\LL(\frac{\mu^2-(\eta-1)^2}{(\eta+{\mu})^2-1}\RR)
  \Bigg]
  +\\&&
  -\frac{g^2}{2\pi m_g}
  \LL(\frac{4\mu^3+2\mu^2+{\mu}}{\mu}\RR)
\end{eqnarray*}
The non-planar parts are
\begin{eqnarray*}
  \frac{\Sigma_{\psi,\rm np}}{m_g}
  &={}&
  \frac{g^2}{2\pi m_g}
   \int \dd x\,
  \Bigg\{
  \LL[
  \frac{1}{\xi}
  -\frac{x\mu^2- x(1-x)\eta^2+x\mu}{\sqrt{x\mu^2-x(1-x)\eta^2+(1-x)}}
  \RR]
  e^{-\xi\sqrt{x\mu^2-x(1-x)\eta^2+(1-x)}}
  +\\&&
  -\LL[
  \frac{1}{\xi}-\frac{x\mu^2-x(1-x)\eta^2+x\mu  }{ \sqrt{x\mu^2-x(1-x)\eta^2} }
  \RR]
  e^{-\xi \sqrt{x\mu^2-x(1-x)\eta^2} }
  \Bigg\}
\\
  \frac{\Sigma_{m,\rm np}}{m_g}
  &={}&
  \frac{g^2}{2\pi m_g}
   \int \dd x\,
  \Bigg\{
  \LL[
  \frac{2(1+\mu)}{\xi}-
  \frac{x\mu^2(1+\mu) +x\eta^2(2x(1+\mu)-2-\mu) }{ \sqrt{x\mu^2-x(1-x)\eta^2} }
  \RR]
  e^{-\xi \sqrt{x\mu^2-x(1-x)\eta^2} }
  +\\&&
  -\LL[
  \frac{2(1+\mu)}{\xi}
  -\frac{ 1+\mu+x(\mu-1)(\mu+1)^2 +x\eta^2(2x(1+\mu)-2-\mu)  }{\sqrt{x\mu^2-x(1-x)\eta^2+(1-x)}}
  \RR]
  \times\\&& \times
  e^{-\xi\sqrt{x\mu^2-x(1-x)\eta^2+(1-x)}}
  \Bigg\}
  .
\end{eqnarray*}
This self-energy is similar to the gauge boson's one, since it is affected by a potentially
unitarity-violating threshold at $p^2=m_f^2$ in addition to the physical one at $p^2=(m_f+m_g)^2$.
We will discuss these terms in the next section, and in section~\ref{sec:Pinch}.

\section{Dispersion relations}\label{sec:dispersion_relations}

The spectrum of the non-commutative Yang--Mills--Chern--Simons system
is entirely encoded in the poles of the propagator\footnote{
  Similar investigations have been performed in~\cite{Barbon:2001dw,Landsteiner:2001ky,Mariz:2006kp}.}.
Firstly, it contains an unphysical pole at $p^2=0$, which
describes the longitudinal degree of freedom still propagating in
any covariant gauge. Secondly, it contains the relevant physical
pole at
\begin{equation}
\label{pole1}
  p^2
  =
  \LL[ m_g^{\rm R}(p,\tilde  p)\RR]^2= \frac{m_g^2 \mathcal{Z}_m^2(p,\tilde p)}{\mathcal{Z}_1(p,\tilde p)
\mathcal{Z}_2(p,\tilde p)},
\end{equation}
which represents the effect of radiative corrections on the
tree-level pole at $p^2=m_g^2$. Since Lorentz invariance is broken,
(\ref{pole1}) does not depend only on $p^2$ but also on  $\tilde
p^2$.

\begin{figure}[htbp]
  \begin{center}
    \includegraphics[width=3in]{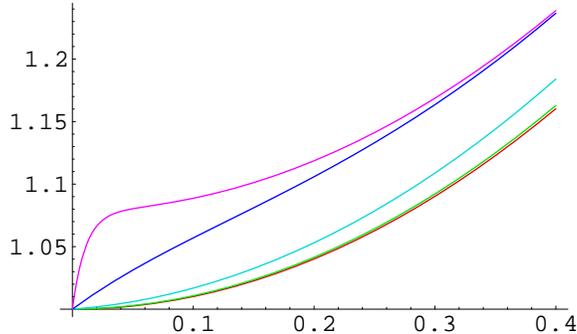}
  \end{center}
  \caption{\label{fig:SUSYDisp}\linespread{0.9}\linespread{0.9}\small{\sf
    Dispersion relations  $E_g^2(|\vec{p}|)$ for both the gauge boson and the gaugino,
    in the fully supersymmetric theory $\mu=1$, for $g^2/m_g=0.1$ and several
    values of non-commutativity $m_g^2\theta\in[10^{-2},10^2]$.
    The dispersion relations are qualitatively similar to the ones displayed here
    for a wide range of parameters~$(g^2,\theta,m_f,m_g)$.
    Both axes are measured in unis of $m_g$.
    }}
\end{figure}

The simplest way to solve  (\ref{pole1}) is to proceed perturbatively\footnote{
  The effects of working with the full (\ref{pole1})  will be discussed in section~\ref{sec:futility}.
}. At the lowest order in $g^2/m_g$ we have the gauge boson's and
fermion's dispersion relations\footnote{In a relativistic theory the
dispersion relations are fixed by Poincar\'e symmetry in the form
$E^2(\vec p)=|\vec p|^2+m^2$.
Non-commutativity however, breaking Lorentz invariance, allows for
non-trivial dependencies of the form $E^2(\vec p)=|\vec
p|^2+m^2+\delta m^2(\vec p)$ --- see figure~\ref{fig:SUSYDisp}, for
example.}
\begin{eqnarray}
  \label{disper1}
  E_g^2
  &=&
  |\vec{p}|^2 +  m_g^2 +\Delta m_g^2(\vec p)
  =
  |\vec{p}|^2+
  m_g^2\Big(
    1
    +\Pi^e_1(p,\tilde p)
    +\Pi^e_2(p,\tilde p)
    -2\Pi^o(p,\tilde p)
  \Big)_{p^2=m_g^2}
  ,
\\
  E_f^2
  &=&
  |\vec{p}|^2 +  m_f^2 +\Delta m_f^2(\vec p)
  =
  |\vec{p}|^2+
  m_f^2\left(
    1+2 \, \frac{\LL(\Sigma_\psi(p,\tilde p)-\Sigma_m(p,\tilde p)\RR)}{m_g}
  \right)_{p^2=m_f^2}
   .
\end{eqnarray}
In order that (\ref{disper1})
provide a reasonable dispersion relation for a stable physical
excitation, two criteria must be met: (a) it has to be gauge
invariant; and (b) it has to be real. These two requirements are far
from being manifest, since the explicit off-shell form of  the $\Pi$'s is
plagued by many complex contributions
coming from the unphysical thresholds at $p^2=0$ and $p^2=m_g^2$, and moreover
our perturbative computation has been performed in the Landau gauge.

However these issues  can be easily clarified by evaluating the
combination $2\Pi^{\rm o}(p,\tilde p) -\Pi^{\rm e}_1(p,\tilde p)-
\Pi^{\rm e}_2(p,\tilde p)$ on shell\footnote{To carry out an
off-shell analysis  the relevant  observable quantity to study is
an $S$-matrix element for a four fermions process. A simpler
approach is provided by the pinch techniques \cite{Caporaso:2005xf}:
in section~\ref{sec:Pinch} we shall sketch how these can be applied
to our case.

} for $p^2=m_g^2$.
A long series of cancellations occur in this quantity, and the final
result is completely real. As we have shown in the previous section
the above combination can be in fact reinterpreted as the
 $S$-matrix element describing a transition between
one-particle states thus, if unitarity is not violated, this element
must be free from unphysical cuts and gauge-invariant.
An analogous mechanism is at work for the fermion's dispersion relation,
and the combination $\Sigma_\psi(p,\tilde p)-\Sigma_m(p,\tilde p)$ is free of spurious thresholds on-shell.

The dispersion relations $E^2(\vec p)=m^2+|\vec p|^2+\Delta m^2(\vec p)$ are given in dimensionless units
\eqref{DimensionlessUnits} as
\begin{align}
     \frac{E^2_{g, f}(\xi;g,\theta,m_f,m_g)}{m_g^2}
     =
  \begin{cases}
     1+\fracT{\xi^2}{(m_g^2\theta)^2}
     +\fracT{\Delta m_g^2}{m_g^2}
     &
     \textrm{gauge bosons},
     \\
     \mu^2
     +\fracT{\xi^2}{(m_g^2\theta)^2}
     +2\fracT{\Delta m_f}{m_g}
     &
     \textrm{fermions},
  \end{cases}
\end{align}
and both can be greatly simplified by performing
some integrations:
\begin{multline}
  \label{deltam_g_soft}
    \frac{ E_{g}^2}{m_g^2}
    =
    1
    +\frac{\xi^2}{(m_g^2\theta)^2}
    +\frac{g^2}{8\pi m_g}
    \LL[
    -(1+2\mu)^2
    \LL(
    \int_{-1/2}^{1/2} \dd x\;
    \frac{1-e^{-\xi\sqrt{x^2+\mu^2-1/4}}}{\sqrt{x^2+\mu^2-1/4}}
    \RR)
    \RR.
    +\\
    +27
    \LL.
    \LL(
    \int_{-1/2}^{1/2} \dd x\;
    \frac{1-e^{-\xi\sqrt{x^2+3/4}}}{\sqrt{x^2+3/4}}
    \RR)
    +4\LL({\mu} -1+\frac{e^{- {\mu} \xi}-e^{-\xi}}{\xi}\RR)
    \RR]
    ,
\end{multline}
and
\begin{multline}
  \label{deltam_f_soft}
    \frac{ E_{f}^2}{m_g^2}
    =
    \mu^2
    +\frac{\xi^2}{(m_g^2\theta)^2}
    +\frac{g^2}{2\pi m_g\mu}
    \Bigg[
    \LL(\frac{e^{-{\mu}\xi}+{\mu}\xi}{\xi}-\frac{e^{-\xi}+\xi}{\xi}
    \RR)
    +\\
    + \frac{(2\mu+1)^2}{2}
    \int_0^1\dd x\,
    \frac{1-e^{-\xi\sqrt{1-x+\mu^2x^2}}}{\sqrt{1-x+\mu^2x^2}}
    \Bigg].
\end{multline}
Both equations decouple for $\xi\rightarrow 0$, as expected, because the group
is U(1)$_\star$ and thanks to supersymmetry. We can distinguish essentially
three terms in the gauge boson's dispersion relation~\eqref{deltam_g_soft}: the
first parenthesis contains the fermion contribution, and the second one
collects the gauge contributions.
The last term collects the pieces (moved from the two preceding parentheses)
that contribute to a ``relic'' of the IR/UV mixing. This expression is finite
in the infrared region $\xi\to 0$ for finite $\mu$, but it has non-trivial
effects, as we will see.

One might wonder whether non-commutativity breaks supersymmetry, and
one possible check is provided by studying the dispersion
relations in the fully supersymmetric case, $\mu=1$.
For $m_f=m_g$ the dispersion relations dramatically simplify, and
we are left just with
\begin{align}
  \label{deltam_g_susy}
    \frac{ E_{g,\rm SUSY}^2}{m_g^2}
    ={}&
    1
    +\frac{\xi^2}{(m_g^2\theta)^2}
    +\frac{9g^2}{4\pi m_g}
    \LL(
    \log 3
    -
    \int_0^1 \dd x\;
    \frac{e^{-\xi\sqrt{1-x+x^2}}}{\sqrt{1-x+x^2}}
    \RR)
    ,
\\
  \label{deltam_majo_susy}
    \frac{ E_{f,\rm SUSY}^2}{m_g^2}
    ={}&
    1
    +\frac{\xi^2}{(m_g^2\theta)^2}
    +\frac{9 g^2}{4\pi m_g}
    \LL(\log 3 -
    \int_0^1 \dd x\; \frac{e^{-\xi\sqrt{1-x+x^2}}}{\sqrt{1-x+ x^2}} \RR)
    .
\end{align}
Since the two dispersion relations coincide one can conclude that
{SUSY} is not broken by non-commutativity at one loop. The
dispersion relations are shown in figure~\ref{fig:SUSYDisp}. For
weak non-commutativity there is no noticeable change with respect to
the free theory; when non-commutativity becomes sufficiently strong,
however, the dispersion relation develops a ``hump''.

\begin{figure}
  \begin{center}
    \includegraphics[width=2.5in]{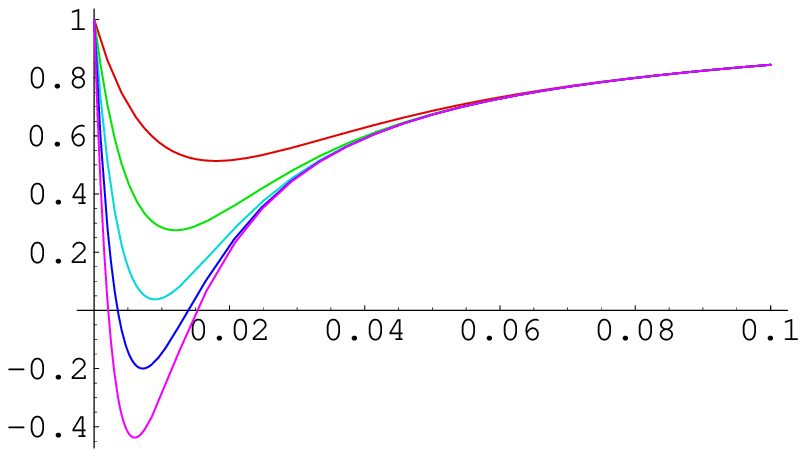}
    \includegraphics[width=2.5in]{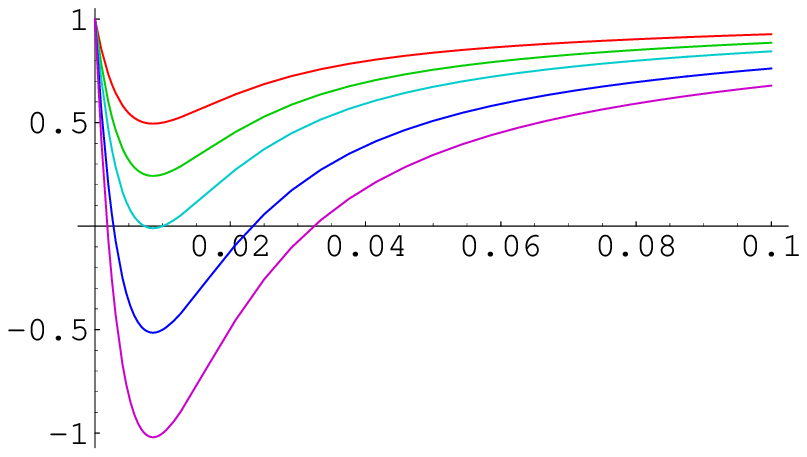}
  \end{center}
  \caption{\label{fig:DispersionRelationDifferentMu}
  \linespread{0.9}\small{\sf
  Tachyonization in the gauge boson's dispersion relation $E_g^2(|\vec{p}|)$ for unit non-commutativity.
  For $|\vec{p}|=0$ we have $E_g^2=1$, as expected from the fact
  that U$(1)_\star$ decouples for $\xi = 0$.
  For $|\vec{p}|>0.1$, the typical behavior $E_g \sim |\vec{p}|^2$
  is recovered, see 
  the following figures.
  Both axes are in units of $m_g$.
  \textit{On the left:}
  small coupling $g^2/m_g=0.1$ and several values of the fermion mass $\mu\in[100,300]$.
  The dispersion relation is anomalous for small values of $|\vec p|$, and becomes
  tachyonic for $\mu$ greater than $\mu_{\rm  crit}\approx 210$.
    \textit{On the right:}
  $\mu=210$, and several values of $g^2/m_g\in[0.05,0.2]$.
    The behavior is qualitatively similar to the one obtaining by varying $\mu$,
    but here the position of the minimum seems to be independent of $g^2/m_g$.
  }}
\end{figure}

\begin{figure}
  \begin{center}
    \includegraphics[width=3in]{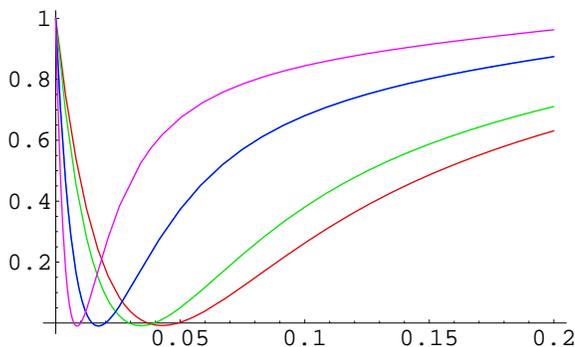}
  \end{center}
  \caption{\linespread{0.9}\linespread{0.9}\small{\sf
    \label{fig:DispersionRelationDifferentz2}
    Marginally tachyonic gauge boson dispersion relation  $E_g^2(|\vec{p}|)$
    for $\mu=210$ and weak coupling $g^2/m_g=0.1$,
    and for several different values of $m_g^2\theta\in[0.2,1]$.
    The qualitative behavior is different from the one of the previous graphs:
    here non-commutativity affects the position of the minimum,
    but one cannot make the particle tachyonic even if
    non-commutativity is taken to be very strong: the mass (or coupling) would have
    to be increased instead.
    Both axes are in units of $m_g$.
  }}
\end{figure}

The gauge boson's dispersion relation in the softly broken case is displayed in
figures~\ref{fig:DispersionRelationDifferentMu}
and~\ref{fig:DispersionRelationDifferentz2}. Some features are evident. The
decoupling is apparent from the fact that the gauge boson's energy $E^2$ tends
to its ``bare'' value $E^2/m_g^2=1$. A striking fact is that the dispersion
relations become anomalous for a certain range of parameters, that is the
mechanical momentum
\begin{align*}
  \vec P\doteq \frac{\dd E}{\dd |\vec p|}\,\hat p
  ,
\end{align*}
and the velocity of the particle are anti-parallel for some values of $|\vec
p|$. Finally, the particle becomes tachyonic in two regions: for large enough
$\mu$, and for $\mu\gtrsim  1/2$ and strong non-commutativity. In the next
sections we will analyze in detail these regions.

The fermion's dispersion relation, on the other hand, is always monotonically increasing
in $|\vec{p}|$ for all the values of the parameters, and so it is
not particularly interesting to study. In the following sections
we concentrate on the gauge boson's dispersion relation,
which as we anticipated has a rather rich behavior.

\subsection{Pure Yang--Mills--Chern--Simons and QCD}

\begin{figure}[htbp]
  \begin{center}
    \includegraphics[width=2.5in]{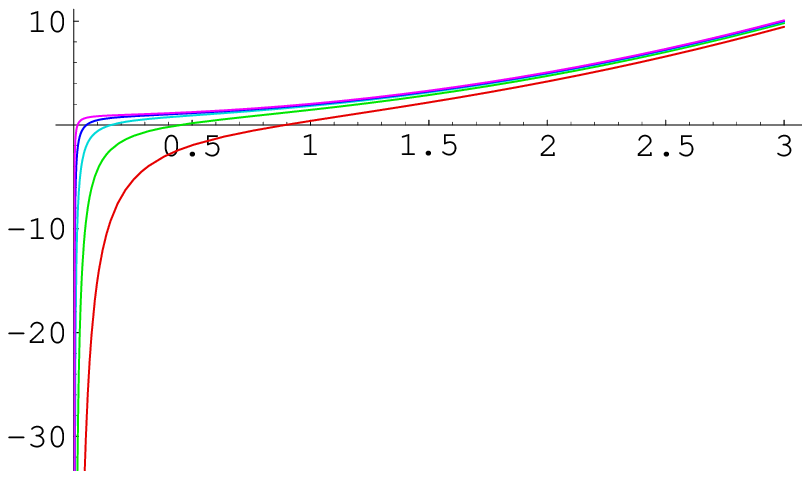}
    \includegraphics[width=2.5in]{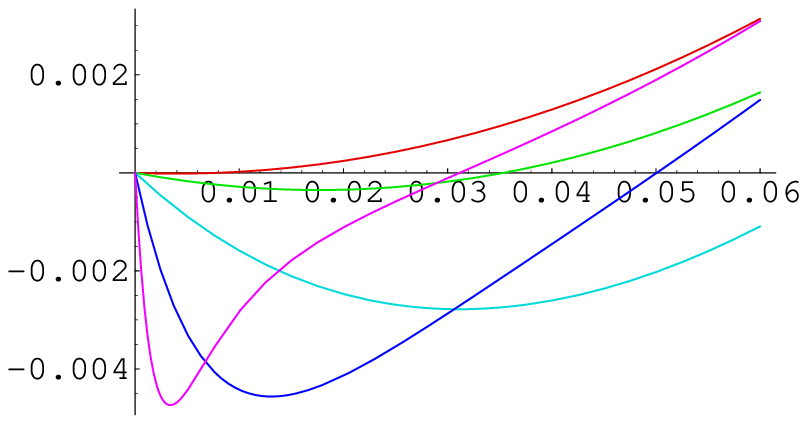}
  \end{center}
  \caption{\linespread{0.9}\small{\sf
      Gauge boson dispersion relations $E_g^2(|\vec{p}|)$ in the $\mu\to\infty$ limit,
      for weak coupling $g^2/m_g=0.1$.
      \textit{On the left:} pure NC-Yang--Mills--Chern--Simons theory, with fermions
      removed, for $m_g^2\theta\in[10,1000]$.
      \textit{On the right}: NC-QCD, for different values of $m_f^2\theta\in[1,100]$.
      In both cases the particles are tachyonic for small values of $|\vec{p}|$, but
      NC-QCD does not have infrared singularities.
      The axes are in units of $m_g$ on the left, and $m_f$ on the right.
      }}
  \label{fig:YMCS.eps}
\end{figure}

As $\mu$ is increased to infinity, two interesting limits are reached:
U(1)$_\star$ YMCS corresponds to $m_f\rightarrow \infty$ with $m_g$ held fixed,
and U(1)$_\star$ QCD corresponds to $m_g\rightarrow 0$ with $m_f$ kept
constant. The way in which these two regions  are approached can be observed in
figures \ref{fig:DispersionRelationDifferentMu} and
\ref{fig:DispersionRelationDifferentz2} while  the two limiting theories are
displayed in figure~\ref{fig:YMCS.eps}.

As one approaches the pure Yang--Mills--Chern--Simons system, the
IR/UV mixing is restored. In fact, an infrared-divergent term of the
form $-4 e^{-\xi}/\xi$ is produced in this limit. The growth of this
negative divergent contribution at small $\xi$ for   sufficiently
large $\mu$ will always make the square of the energy negative in a
certain region of spatial momenta (see the figure). In other words,
the massive excitation becomes a tachyon and the perturbative vacuum
is no longer stable. Varying the other two parameters $(g^2,\theta)$
will not affect the picture: it will only change the specific value
of $\mu$ at which the tachyon will appear. Thus, when we reach the
critical value of $\mu$, we must  resort to non-perturbative
tecniques to select the new vacuum. The gauge boson's dispersion
relation for $m_f\rightarrow\infty$ reads
\begin{align}
  \begin{split}\label{deltam_decoupling}
    \frac{ E_{g,(m_f\rightarrow\infty)}^2}{m_g^2}
    ={}&
    1
    +\frac{\xi^2}{(m_g^2\theta)^2}
    +\frac{g^2}{8\pi m_g}
    \Bigg\{
    -27
    \LL(
    \int_0^1 \dd x\;
    \frac{e^{-\xi\sqrt{1-x+x^2}}-1}{\sqrt{1-x+x^2}}
    \RR)
    -4\LL(\frac{e^{-\xi}+\xi}{\xi}\RR)
    -4
    \Bigg\}
    .
  \end{split}
\end{align}
The last factor of $-4$ in the gauge boson's dispersion relation
is essentially the parity-violating anomaly from the decoupled fermion loop.
We see that an IR pole has appeared, as shown in figure~\ref{fig:YMCS.eps}.
This is a nice signature of the reappearance of the infrared divergences
when the regulator $\Lambda=m_f$ is removed.
It is in fact a well-known problem in non-commutative field theories
that the physics depends on the order in which the
limits $\theta\rightarrow 0$ and $\Lambda\rightarrow\infty$
are taken.

In the limit $m_g\rightarrow 0$, where non-commutative QCD is reached, we can no
longer use $m_g$ as a meter, so we divide
equation~\eqref{deltam_g_soft} by $\mu^2$ and equation
\eqref{deltam_f_soft}  by $\mu$ to re-express everything in units of $m_f$.
Instead of $\xi$, which includes a factor of $m_g$, only in this section
we will work with
\(
     \zeta  \doteq{} m_f \theta |\vec p|
\).
In the limit ${\mu}\rightarrow\infty$,
in which $\zeta$ and $g^2/m_f$ stay finite, the dispersion relations become
\begin{align}
  \begin{split}\label{deltam_mg=0}
    \frac{ E_{g,\rm QCD}^2}{m_f^2}
    ={}&
    \frac{\zeta^2}{m_f^4\theta^2}
    +\frac{g^2}{2\pi {m_f}}
    \LL(
    e^{-\zeta}+\frac{e^{-\zeta}-1}{\zeta}
    \RR)
=
    \frac{|\vec{p}|^2}{m_f^2}
    +\frac{g^2}{2\pi m_f}
    \LL(
    e^{-{m_f}\theta |\vec{p}|}+\frac{e^{-{m_f}\theta |\vec{p}|}-1}{{m_f}\theta |\vec{p}|}
    \RR)
    ,
\\
    \frac{ E_{f,\rm QCD}^2}{m_f^2}
    ={}&
    1
    +\frac{\zeta^2}{m_f^4\theta^2}
    +\frac{g^2}{2\pi m_f}
    \LL[
    \LL(\frac{e^{-\zeta}-(1-\zeta)}{\zeta}\RR)
    +2\LL(\gamma+\Gamma[0,\zeta]+\log \zeta\RR)
    \RR]
    .
  \end{split}
\end{align}
In this case, the gluon has a highly non-trivial dispersion relation,
characterized by a tachyonic effective mass for small values of $p$, as can be seen in figure~\ref{fig:YMCS.eps}.
Now no particle has been removed from the spectrum, so that no IR singularity is generated.
Nevertheless, expanding the dispersion
relation for small values of $p$, one immediately sees that tachyonic modes develop also in this limit.
An IR singularity is restored by taking $m_f^2\theta\rightarrow\infty$.

In the quark's dispersion relation, $\gamma$ is Euler--Mascheroni's
constant and $\Gamma[0,\zeta]$ is an incomplete Gamma
function. The Gamma function and the logarithm separately diverge
for $\zeta \rightarrow 0$, but their sum does not.
The appearance of the Euler--Mascheroni constant in an observable quantity is an exotic feature:
in commutative quantum field theories it is a well-known
fact that this constant always cancels in the
observable quantities\footnote{See, for example, the book by Peskin and Schroeder \cite{Peskin:1995ev}.}.
We will provide an interpretation for this finding in
section~\ref{sec:why_they_occur}.

\subsection{Near-threshold dispersion relation}

\begin{figure}
  \begin{center}
    \includegraphics[width=3in]{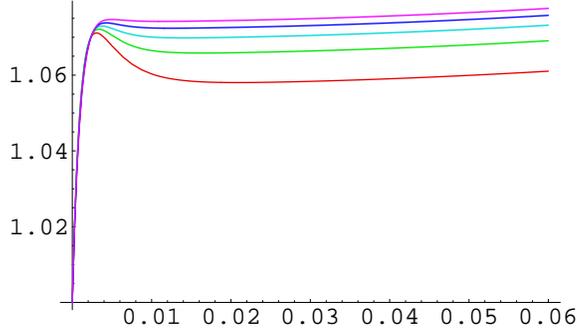}
  \end{center}
  \caption{\label{SoftMassShiftNearThreshold1}\linespread{0.9}\small{\sf
    The dispersion relation becomes  $E_g^2(|\vec{p}|)$ anomalous as $\mu$ approaches the threshold $\mu=1/2$.
    Here $g^2/m_g=0.1$, $m_g^2\theta=10^3$ and $\mu\in[0.55,0.75]$.
    This phenomenon is however completely different from the large $\mu$ one,
    since for small enough values of $|\vec p|$ here the dispersion relation
    is always non-anomalous.
    Both axes are in units of $m_g$.
  }}
\end{figure}

\begin{figure}
  \begin{center}
    \includegraphics[width=3in]{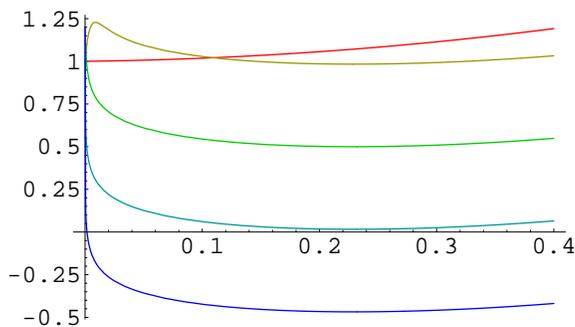}
  \end{center}
  \caption{\label{SoftMassShiftNearThreshold2}\linespread{0.9}\small{\sf
    The dispersion relation $E_g^2(|\vec{p})$ at the threshold $\mu=1/2$ and for $g^2/m_g=1/3$
    becomes anomalous and then tachyonic as non-commutativity is increased,
    at $m_g^2\theta=3\times 10^6$.
    The range here is $m_g^2\theta=[0.3,0.3\times 10^9]$.
    Notice that the position of the minimum is independent of $\theta$.
    The non-anomalous region of small momenta becomes progressively obscured,
    but a numerical analysis reveals that a tiny non-anomalous region continues to exist.
  }}
\end{figure}

We now come to the dispersion relation for the gauge boson in the case $\mu\gtrsim 1/2$,
where the masses are near the threshold for the decay process $g\to f\bar f$.
In the specific case $\mu=1/2$ one integral is easily computed, and the result
is rather simple
\begin{align}
  \begin{split}\label{deltam_g_threshold}
    \frac{ E_{g,\rm threshold}^2}{m_g^2}
    ={}&
    1
    +\frac{\xi^2}{(m_g^2\theta)^2}
    +\frac{g^2}{8\pi m_g}
    \Bigg\{
    8
    \big[
    -\gamma - \Gamma(0, \xi/2) - \log\LL({\xi}/{2}\RR)
    \big]
    +\\& \phantom{\frac{g^2}{8\pi m_g}\Bigg[}
    -27
    \LL(
    \int_0^1 \dd x\;
    \frac{e^{-\xi\sqrt{1-x+x^2}}-1}{\sqrt{1-x+x^2}}
    \RR)
    +4\LL(-\frac{1}{2}+\frac{e^{- \xi/2}-e^{-\xi}}{\xi}\RR)
    \Bigg\}
    ,
  \end{split}
\end{align}
where $\gamma$ is again Euler-Mascheroni's constant and $\Gamma(0,
\xi/2)$ is an incomplete Gamma function.
Below $\mu=1/2$ imaginary parts start,
as expected from the optical theorem because the gauge boson becomes unstable.
The behavior  of the dispersion
relation as the fermion mass reaches the value $m_f=(1/2) m_g$ from above
is displayed in figure~\ref{SoftMassShiftNearThreshold1}. Figure
\ref{SoftMassShiftNearThreshold2} shows the dispersion relation for
$\mu=1/2$ and for different values of $m_g^2\theta$: the gauge boson
becomes tachyonic for very large values of $m_g^2\theta$.
Considering a particle whose mass sits exactly at the threshold for its decay process
is analytically convenient, but physically rather tricky,
particularly when the particle becomes a tachyon.
Nevertheless, it is not hard to check numerically that
for a stable gauge boson of mass $\mu\gtrsim 1/2$,
the dispersion relation is qualitatively very similar and
the limit $\mu\rightarrow 1/2$ is smooth.

A few remarks on the dispersion relations in the $\mu\gg 1$ and $\mu \gtrsim
1/2$ cases are useful at this point. The physics in the two ranges of
parameters looks similar at a first glance: in both cases, in fact, the
dispersion relation becomes anomalous for a finite value of $|\vec p|$, and
then tachyonic as some parameter is increased. On a closer scrutiny, however,
one notices several important differences.
\begin{itemize}
  \item  The most evident one is probably that while
  for $\mu\gg 1$ the dispersion relation is always anomalous
  for small enough values of $|\vec p|$ (see
  figures~\ref{fig:DispersionRelationDifferentMu} and
  \ref{fig:DispersionRelationDifferentz2}), in the $\mu\gtrsim  1/2$
  case there seems to be a finite neighbourhood of $|\vec p|=0$
  where the dispersion relation is non-anomalous, as one sees rather
  clearly in figure~\ref{SoftMassShiftNearThreshold1}.

  \item  In addition, in the $\mu\gg 1$ case the gauge boson becomes
  tachyonic for large enough values of $\mu$, even if $m_g^2\theta$ is
  not very large. The opposite is true in the $\mu \gtrsim  1/2$ region, where
  very strong non-commutativity is needed to generate tachyonic
  modes, see figure~\ref{SoftMassShiftNearThreshold2}.

  \item  Finally,
  in the large $\mu$ region one can see from figure
  \ref{fig:DispersionRelationDifferentz2} that the position of the
  minimum depends on both the gaugino mass $\mu$ and on the
  non-commutativity parameter $m_g^2\theta$ but not on $g^2/m_g$.
  In the threshold region, on the other hand,
  $m_g^2\theta$ does not affect the position of the minimum of
  the energy, as is evident from figure
  \ref{SoftMassShiftNearThreshold2}: the position of the minimum is
  determined solely by the coupling constant $g^2/m_g$.
\end{itemize}
These findings are worthy of being analyzed in more detail,
and in the next section we will see that there are two very
different mechanism at work behind the two instability regions encountered
in the $\mu\gg 1$ and $\mu\gtrsim 1/2$ cases.

\section{Why (and how) anomalous dispersion relations and tachyonization occur}\label{sec:why_they_occur}

To explain how anomalous dispersion relations and tachyonization occur,
we need a clearer picture of the structure of the one-loop quantum
effects in this theory.
As we have shown, after all the terms are summed up one finds that in the IR,
which corresponds to $\theta |\vec{p}|\doteq\sqrt{p\bullet p}=\theta |\vec{p}|=0$, there are no quantum effects
at all, thanks to a perfect cancellation among all the contributions.
So we see that at $\theta |\vec{p}|=0$ we have a free theory.
As $\theta |\vec{p}|$ is increased from zero, all the interactions  are turned on
in a rather complicated fashion.
Finally, as we take $\theta |\vec{p}|\rightarrow \infty$, all non-planar diagrams are
exponentially suppressed:
in this limit the theory turns into commutative topologically-massive QED\footnote{
  For this picture to be meaningful one must substitute Dirac to Majorana fermions of
  course, since Majorana fermions are non-interacting in the commutative theory.
}.
From this point of view, we can see the parameter $\theta |\vec{p}|$ as an effective cut-off  $\Lambda_{\rm eff}=\theta p$ which turns on the interactions smoothly as it interpolates between a free theory (at $\theta |\vec{p}|=0$) and a commutative theory (at $\theta |\vec{p}|=\infty$).

\begin{figure}[htbp]
  \begin{center}
    \includegraphics[width=1.9in]{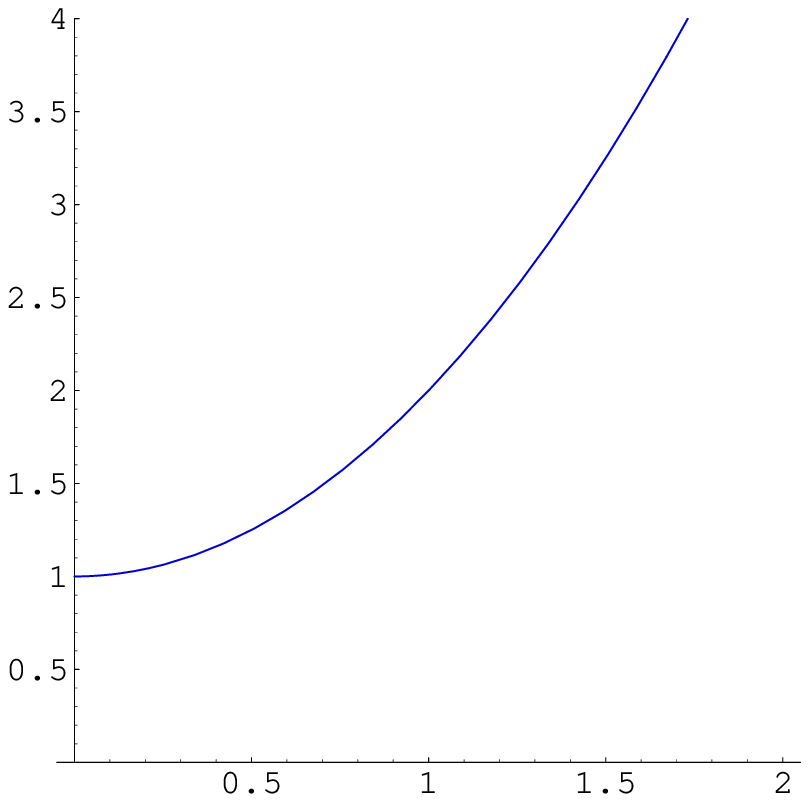}
    \includegraphics[width=1.9in]{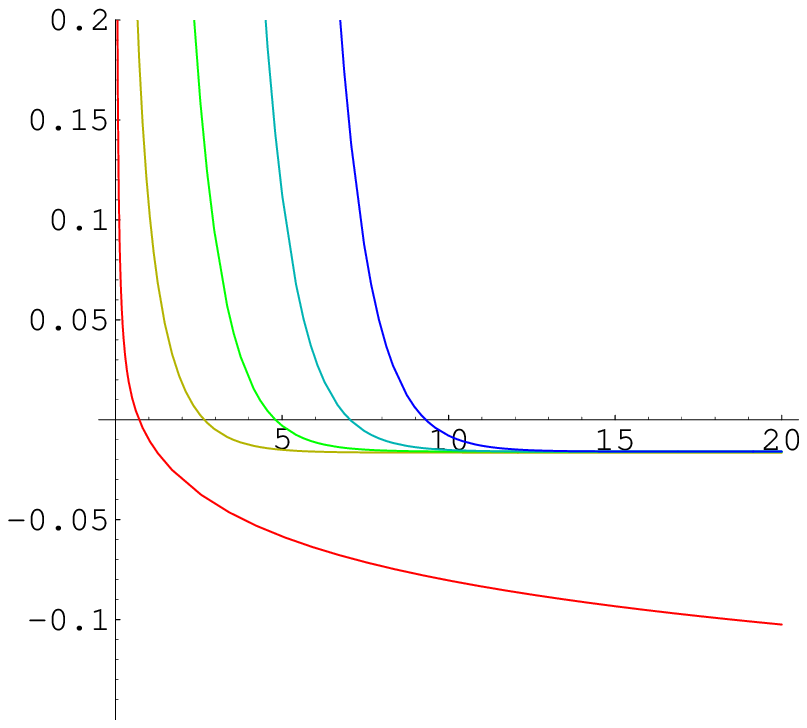}
    \includegraphics[width=1.9in]{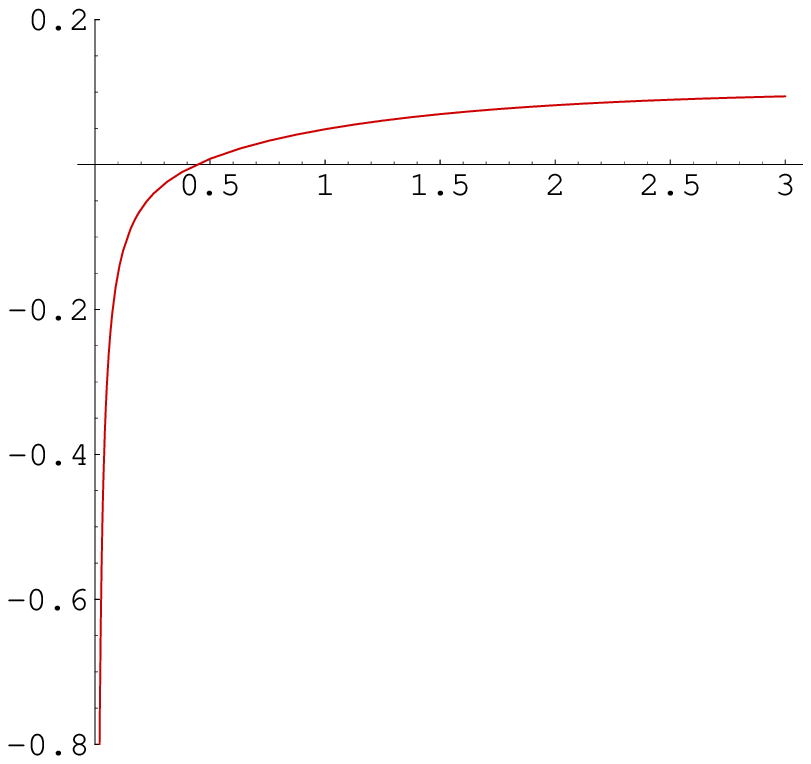}
  \end{center}
  \caption{\linespread{0.9}\small{\sf The three contributions to the dispersion relation, from left to right:
  (i) the tree-level dispersion relation, as a function of $|\vec{p}|/m_g$;
  (ii) the repulsive fermion loop contribution to the dispersion relation, as a function of
       $\zeta=m_f\theta|\vec{p}|$, for $\mu=\{0.5,10^1,10^2,10^3,10^4\}$;
  (iii) the attractive gauge sector contribution as a function of $\xi=m_g\theta|\vec{p}|$.
  }}
  \label{ThreeContributions.eps}
\end{figure}

By inspecting the contributions to the polarization tensor in section~\ref{sec:polarization_tensor},
however, one finds several interesting common features.
For example, the fermionic contributions $\Pi^{\rm ferm,e}_1$, $\Pi^{\rm ferm,e}_2$ and  $\Pi^{\rm o,ferm}$
are monotonically decreasing in $\theta |\vec p|$ when evaluated on-shell.
In addition, the strength of these contributions is damped by the factors of $\mu$ and $\xi$
appearing in the exponentials.
This means that the variations in $p$ of the fermionic contributions
are governed by a ``characteristic length" $\ell_f$:
\begin{align}
  \text{fermion loop:}
  \qquad
  \label{ell_eff}
  \exp(-p\ell_f),
  \qquad
  \ell_f
  =
  \theta\sqrt{m_f^2-x(1-x)m_g^2}
  \,\approx\,
  \theta \cdot
  \max(m_f,m_g)
  .
\end{align}
Similarly, one can show that the whole contribution from the gauge loops is always
monotonically increasing with $p$.
In this case,
the exponential factors are damped by different characteristic lengths,
all of which are  on the order of $m_g\theta$, so we can roughly say
\begin{align}
  \label{ell_gee}
  \text{gauge loops:}
  \qquad
  \exp(-p\ell_g),
  \qquad
  \ell_g
  =
  \theta\sqrt{f(x)}
  \approx m_g\theta.
\end{align}
The fermionic and gauge contributions, along with the standard ``kinetic
term'', are depicted in figure~\ref{ThreeContributions.eps}. Notice the
different scales on the horizontal axis: $|\vec{p}|/m_g$, $m_f\theta|\vec p|$,
and $m_g\theta|\vec p|$, and the infrared divergences in the second and third
graphs. The second graph also displays an UV logarithmic divergence for
$\mu=1/2$, which will be discussed in section~\ref{sec:WhyThresh}.

\subsection{Tachyonization for large $\mu$: infrared/ultraviolet mixing}

The main reason behind the occurrence of anomalous
dispersion relations and tachyonization for large $\mu$
is that, as we have seen,
relatively heavy fermions influence the small $p$
physics sooner than the gauge bosons do ---
see equations~\eqref{ell_eff} and~\eqref{ell_gee}.
So we expect, for heavy enough fermions, a region of small
values of $|\vec p|$ where the dispersion relation is anomalous.

Let us see how this works in a more quantitative way.
We expand the single contributions to ${\rm O}(\xi)$ and find
\begin{align}
  \begin{split}
    \frac{(\Delta m_g^{\rm ferm})^2}{m_g^2}
    ={}&
    \frac{g^2}{8\pi m_g}
    \LL(
    \frac{4}{\xi}
      -\LL(2\mu^2+4\mu+1\RR) \xi
      +{\rm O}(\xi^2).
    \RR)
\\
    \frac{(\Delta m_g^{\rm glue})^2}{m_g^2}
    ={}&
    \frac{g^2}{8\pi m_g}
    \LL(
      -\frac{4}{\xi}+25\xi
      +{\rm O}(\xi^2),
    \RR)
  \end{split}
\end{align}
notice that, again, the fermion contribution is always decreasing
and the bosonic contribution has the opposite behaviour.
The grand total is
\begin{align}
      \frac{E_g^2}{m_g^2}
      =
      1+\frac{|\vec p|^2}{m_g^2}
      +
    \frac{g^2}{4\pi m_g}
    \LL(
    -\mu^2-2\mu+12
    \RR)
    m_g\theta|\vec p|+{\rm O}(g^2\theta|\vec p|)^2
    .
\end{align}
The group velocity is defined by $v_g \doteq \frac{\dd E_g}{\dd
|\vec{p}|}$  and for  $|\vec{p}|=0$ one finds
\begin{align}
  v_g\Big|_{|\vec{p}|=0}
  =
  \frac{m_g^2\theta}{4\pi}
  \frac{g^2}{m_g}
  [12-\mu(2+\mu)]
  .
\end{align}
We see that the dispersion relation becomes anomalous for
\begin{align}\label{CritAnomal}
  \mu
  >
  \sqrt{13}-1
  ,
\end{align}
numerical checks confirm this computation. Because of the anomalous region,
there exists a nonzero characteristic momentum $|\vec p|=\bar p$ for which the
particle's squared energy has a minimum. By increasing $\mu$ further, this
value can become negative, leading to a tachyonic dispersion relation.

For tachyonization to occur at weak coupling,
one needs to have large contributions from the quantum
loops\footnote{These large coefficients do not lead
  the theory away from the domain of perturbation theory, as we will see in section~\ref{sec:HigherOrders}.
}, which must overcome the factors of $g^2$. The only possibility is
that these contributions originate from the IR/UV phenomenon,
whenever the fermionic and bosonic ``characteristic lengths"
$\ell_i$ are of different orders of magnitude. If this is the case,
tachyonization should occur because even though at $|\vec{p}|=0$
there is a perfect cancellation between the IR divergences, for some
small value of $|\vec{p}|$ large shifts in the gauge boson's
renormalized mass appear. The scale where this happens should be
determined by the fermion ``characteristic length" $\ell_f\sim\theta
m_f$. Infact in  the regime where $|\vec{p}|\ell_f\sim 1$ and
$|\vec{p}|\ell_g\ll 1$ the non-planar diagrams which contain fermion
loops are  dumped and  the negative bosonic contributions prevail.
More specifically, we expect the tachyon to have a characteristic
momentum of order
  \begin{align} \label{forecast_infty_form}
    \bar{p}
     \stackrel?\propto
    \frac{1}{\theta m_f}
    .
  \end{align}
We shall now prove this hypothesis.

To test this picture, we need to find for what critical values of the parameters $(g,\theta,m_g,\mu)$
the dispersion relation first develops a massless mode, which will become tachyonic when the critical line is crossed.
Looking at figure~\ref{fig:DispersionRelationDifferentMu},
we see that we must first of all identify the characteristic momentum $\bar{p}(g,\theta,m_g,\mu)$
\begin{equation}
  \label{find_tachyon_1}
  \bar p(\mu,g,\theta,m_g):
  \qquad
  \left.\frac{\dd E_g^2\big(|\vec{p}|;\mu,g,\theta,m_g\big)}{\dd|\vec{p}|}
  \right|_{\bar{p}=|\vec{p}|}
  =0
  ,
\end{equation}
and then find for what value $\mu_{\rm crit}(g,\theta,m_g)$
of the fermion's mass the energy $E^2(\bar{p})$ of this mode equals zero:
\begin{equation}
  \label{find_tachyon_2}
  \mu_{\rm crit}(g,\theta,m_g):
  \qquad
  E_g^2\big(\bar{p}(\mu_{\rm crit},g,\theta,m_g);\mu_{\rm crit},g,\theta,m_g\big)
  =
  0
  .
\end{equation}
Along this critical line of parameters, the dispersion relation will have a
massless mode $\bar p$, like in figure~\ref{fig:DispersionRelationDifferentz2}.
In the following, it will be more practical to look for the minimum of the dispersion relation
in the related variable  $\bar\xi = m_g\theta \bar p$.

\begin{figure}[htbp]
    \begin{center}
      \includegraphics[width=3in]{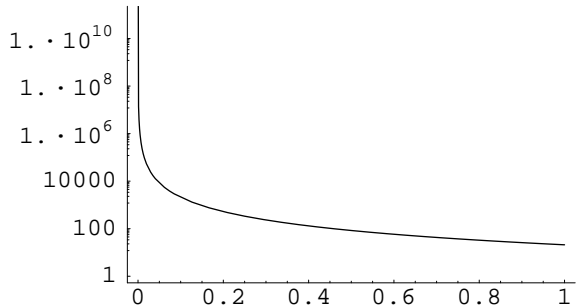}
    \end{center}
    \caption{\linespread{0.9}\small{\sf
      Destabilization line $\mu_\text{crit}(g^2/m_g)$.
      Above the destabilization line, a tachyonic mode exists.
      This is valid unless $m_g^2\theta\ll (\mu)^{-1}$.
     }}
      \label{heavy_fermion.eps}
  \end{figure}

Unfortunately, even in the large $\mu$ limit equation~\eqref{find_tachyon_1} turns out to be quite complicated:
taking the derivative of~\eqref{deltam_g_soft} with respect to $\bar\xi=m_g\theta\bar p$
one gets an equation that cannot be solved analytically.
Clearly, however, in the large $\mu$ limit a few of the terms of~\eqref{deltam_g_soft} will be neglegible.
One can test several different large-$\mu$ behaviors for
$\bar \xi$, and it is not hard to convince oneself that only
\begin{align}
  \label{LargeMuHyp}
    \bar\xi =  \frac{\beta(\mu,g,\theta,m_g)}{\mu}
    ,
\end{align}
gives a non-trivial result for equation~\eqref{find_tachyon_1} ---
here, $\beta(\mu,g,\theta,m_g)$ is a constant to be determined using
equation~\eqref{find_tachyon_1}.
In this case, in fact,
there is a competition between several addenda in the dispersion relation,
and one obtains
\begin{align}
    \label{critDR}
    \frac{ E_{g,\mu\rightarrow\infty}^2}{m_g^2}
   ={}&
    1
    + \frac{g^2}{2\pi m_g} \left[
      (e^{-\beta}-1)
      + \mu\left( e^{-\beta }+\frac{e^{-\beta}-1}{\beta }\right)
    \right]
   +\textstyle{{\rm O}\left(\frac{g^2}{m_g \mu},\frac{1}{m_g^4\theta^2\mu^2}\right)}
    .
\end{align}
which does have a non-trivial minimum.
Therefore, the constant $\beta$ in equation \eqref{LargeMuHyp}
is to be determined as the minimum of the quantity~\eqref{critDR}:
\begin{align}
   1+\beta + \beta ^2
   - e^{\beta }
   =
   - \frac{\beta^2}{\mu}
   ,
\end{align}
which, for large $\mu$, has the non-trivial solution $\beta\approx 1.8$.
The first tachyon mode is generated when the dispersion relation~\eqref{critDR},
evaluated in the ``softest'' momentum $\beta\approx 1.8$, is equal to zero.
This happens for
\begin{equation}\label{large_mu_1}
  \mu_{\rm crit}
   \approx
   \frac{21 m_g}{g^2}-2.8
   +\textstyle{{\rm O}\left(\frac{g^2}{m_g \mu},\frac{1}{m_g^4\theta^2\mu^2}\right)}
\end{equation}
The destabilization line is displayed
in figure~\ref{heavy_fermion.eps} for several values of the
coupling.
We see that, by taking $\mu$ to be large enough, the transition
always occurs for $g^2/m_g\ll 1$.
This is important when one considers the effect of higher orders
in perturbation theory, see section~\ref{sec:HigherOrders}.

The fact that  we obtain a finite value for $\beta$ confirms that we are really
sitting on the dispersion relation's minimum, and that $\bar\xi\sim\mu^{-1}$
is the correct large-$\mu$ behaviour for the momentum $\bar p$.
This confirms $\bar p$ to be
of the order of the ``characteristic length" $\ell_f$, as we anticipated
in~\eqref{forecast_infty_form}.
The phenomenon underlying the generation of these tachyonic modes is indeed
an ``imperfect'' cancellation of the IR/UV divergence\footnote{
  In deriving the dispersion relation~\eqref{critDR}, we have neglected the kinetic energy,
  a contribution of order $(m_g^2\theta\mu)^{-2}$.
  Clearly, the transition line~\eqref{large_mu_1} will be independent of $\theta$
  as long as $m_g^2\theta\geq {\rm O}(\mu^{-1})$.
  It is in fact possible to take two more orders in $\mu$ in equation~\eqref{critDR},
  and solve the resulting destabilization equation to get either
  $\mu_{\rm crit}(g,\theta)$ or $\theta^{\rm crit}(g,\mu)$
  exactly, up to ${\rm O}(\mu^{-3})$.
  The result is rather complicated and uninstructive (it is a third-order equation in $\mu$),
  and confirms that $\mu_{\rm crit}(g,\theta)\approx\mu_{\rm crit}(g)$
  unless one takes extremely small values of $m_g^2\theta$.
  The fact that the kinetic energy only contributes a higher-order correction
  is a signature of the fact that the interplay between the IR-singular terms
  dominates.
}.

\subsection{Tachyonization for $\mu\gtrsim 1/2$:  ultraviolet/mass renormalization mixing}\label{sec:WhyThresh}

Let us now turn to the second anomalous region, which corresponds
to $\mu \gtrsim  1/2$ and strong non-commutativity.
From the above reasonings and the general lore on perturbative non-commutative
theories, the first guess would be that the tachyonic
instabilities are generated by the IR/UV phenomenon also here.

So far, in fact, we have only found negative contributions to the dispersion relation from the gauge and ghost loops.
These terms cannot be considered responsible in this second case.
The reason is that since $\mu\gtrsim  1/2$,
there is a ``good'' cancellation between the fermionic and bosonic IR singularities,
and the IR/UV phenomenon cannot lead to tachyonization.
In fact, for $|\vec{p}|\approx 0$ the dispersion relation is not even anomalous,
as one can see in equation~\eqref{CritAnomal} and in
figures~\ref{SoftMassShiftNearThreshold1} and~\ref{SoftMassShiftNearThreshold2}.
This leads to a first puzzling aspect of this destabilization:
we seem to be too ``near'' to the supersymmetric theory to have large terms.
The mechanism by which a tachyonic mass is
generated in this region must be completely different from the one
of the previous section.

\begin{figure}[htbp]
  \begin{center}
    \includegraphics[width=3in]{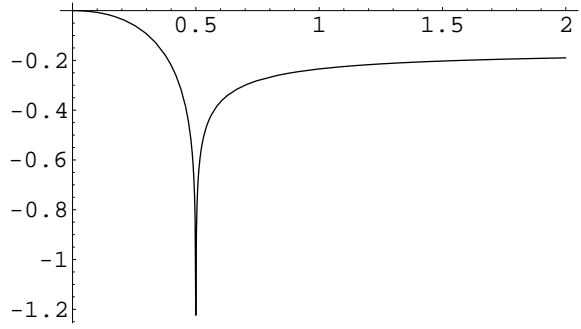}
  \end{center}
  \caption{\linespread{0.9}\small{\sf
  \label{fermion_loop_pla}
  Fermion loop contribution to the mass renormalization of the gauge boson
  $\delta m_g(\mu)$
  in topologically massive QED,
  corresponding in our theory to the planar limit of the Majorana gaugino loop,
  on-shell, as a function of $\mu$. $g^2/m_g=1$.
  The divergence for $\mu=1/2$ is responsible for the new tachyonic modes.
  Below $\mu=1/2$, the real part is plotted.
  }}
\end{figure}

The only possibility is that, in some sense, this second
tachyonic region has its origin for large $\xi$, where the physics
is related to commutative topologically massive QED.
In the large-$\xi$ limit the various contributions tend to their planar limits.
The gauge and ghost planar terms are positive.
On the contrary, the fermionic contribution in section~\ref{sec:polarization_tensor}
is always negative and monotonically decreases from zero to its planar limit, which is
\begin{align}
  \begin{split}
    \LL(\Pi^{\rm e,ferm}_{\rm 1}
    +\Pi^{\rm e,ferm}_{\rm 2}
    -2\Pi^{\rm o,ferm}
    \RR)
    \stackrel{(\theta p\rightarrow\infty)}{\longrightarrow}
    \frac{g^2}{32\pi m_g}
    \LL[
    4\mu
    +\left(1+2\mu \right)^2
    \log \LL(\frac{2{\mu}-1}{2{\mu}+1}\RR) \RR]
    .
  \end{split}
\end{align}
This is displayed in figure~\ref{fermion_loop_pla}.
We notice a divergence for $\mu\to 1/2$: in topologically massive
QED the fermions induce a negative mass shift which diverges
logarithmically at the threshold for the process $g \rightarrow ff$.
Hence for $\mu=1/2$, in the NC theory, the renormalized
gauge boson mass at $\xi=\infty$ is infinitely lower than that at
$\xi=0$, and this should explain the destabilization of the
perturbative vacuum.
The divergence for $\xi\rightarrow\infty$ at $\mu=1/2$ from this contibution is clearly
visible in the second plot in figure~\ref{ThreeContributions.eps}.

Let us explore the possibility for this transition to occur at large $\xi$.
This means that the destabilizing quantum effects must
compete with the kinetic $|\vec p|^2$ term in the dispersion relation.
This can happen if the transition occurs for $m_g^2\theta \gg 1$,
 so that the ``bare'' term $|\vec p|^2=\xi^2/(m_g^4\theta^2)$ is suppressed.
But on the other hand, we have found numerically that the
position of the minimum appears to depend on $g$, but
not on $\theta$ (see, for example, figure~\ref{SoftMassShiftNearThreshold2}). How can this be so?

To investigate what is the effect of having large values of $\xi$, let us
inspect the case $\mu=1/2$ directly
to get an estimate of the point where the transition first occurs:
\begin{align}
  \begin{split}
    \label{Disper1/2}
    \frac{ E_{g}^2}{m_g^2}
    ={}&
    1+\frac{\xi^2}{(m_g^2\theta)^2}
    +\frac{g^2}{8\pi m_g}
    \Bigg[
    -8
    \LL(
    \gamma + \Gamma(0, \xi/2) + \log\LL({\xi}/{2}\RR)
    \RR)
    +\\& \phantom{\frac{g^2}{8\pi m_g}\Bigg[}
    +27
    \LL(
    \int_{-1/2}^{1/2} \dd x\;
    \frac{1-e^{-\xi\sqrt{x^2+3/4}}}{\sqrt{x^2+3/4}}
    \RR)
    +4\LL(-1/2+\frac{e^{-\xi/2}-e^{-\xi}}{\xi}\RR)
    \Bigg]
    .
  \end{split}
\end{align}
$\gamma$ is again Euler--Mascheroni's constant.
In the first row we have (part of) the contribution
from the fermion loop. The second line is always positive, so the
transition must be caused by the terms in the first line,
which gets below $-1$ for $g^2/m_g=1$ only at about
$\xi=30$. This means that, for weak coupling, there is no
transition unless $\xi\gg 1$, as we expected.
The integral in the second row can be approximated for large $\xi$
by taking its planar limit, and the incomplete Gamma function is exponentially
depressed for large $\xi$ as well, so we get
\begin{align*}
    \label{forecast_1/2_form}
    \frac{ E_{g}^2}{m_g^2}
    \stackrel{(\xi\gg 1)}{\approx}{}
    1+\frac{\xi^2}{m_g^4\theta^2}
    +\frac{g^2}{8\pi m_g}
    &\LL(
    27\log 3
    -2
    -8\gamma
    -8 \log\frac{\xi}{2}
    \RR)
    .
\end{align*}
We have seen that the divergence in the
mass renormalization of topologically massive QED as $m_f\rightarrow m_g/2$ is logarithmic.
In the UV ($\theta|\vec p|\rightarrow \infty$) domain, this divergence becomes
a logarithmic divergence in $\xi$.
We can call this phenomenon an \textit{ultraviolet/mass renormalization mixing} (UV/MR).

\begin{figure}
  \begin{center}
    \includegraphics[width=3in]{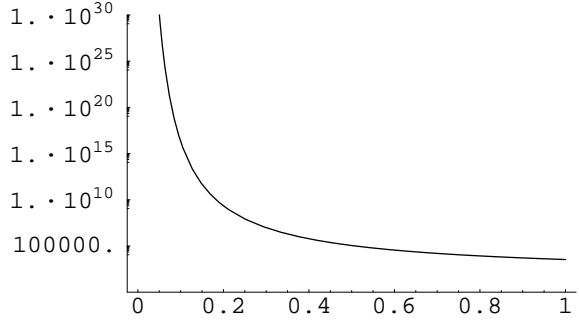}
  \end{center}
  \caption{\label{fig:DestabLineThresh.eps}\linespread{0.9}\small{\sf
  Destabilization line $\theta^{\rm crit}(g)$ for $\mu=1/2$.
  Above this critical line, a tachyonic mode with momentum
  \(
    \bar{p}\approx\sqrt{g^2 m_g/2\pi}
  \)
  is generated.
  Both axes are in units of $m_g$.
  }}
\end{figure}

The need for extremely large values of $m_g^2\theta$ and the
independence of the minimum from $m_g^2\theta$ follow quite
naturally from the fact that the divergence is logarithmic. The
characteristic $\bar p$ is given by
\begin{align}
  \bar\xi
  \approx{}&
  m_g^2\theta\sqrt{\frac{g^2}{2\pi m_g}}
  &\Leftrightarrow&&
  \bar p
  \approx{}&
  \sqrt{\frac{g^2 m_g}{2\pi} }
\end{align}
which is independent of $\theta$, as expected.
Let us now solve the equation $E_g(\bar\xi)=0$.
We find that a gauge boson with momentum $\bar p$ gets tachyonic for
exponentially large values of
\begin{align}
  \label{BigTheta}
    m_g^2\theta^{\rm crit}
    \approx{}&
    2 \, \exp
    \LL\{
    \frac{\pi m_g}{g^2}+\frac{27\log{3}-1}{8}-\gamma
    \RR\}
    \approx
    2 \, e^{3+\pi m_g/g^2}.
\end{align}
This region is shown in figure~\ref{fig:DestabLineThresh.eps}.
Comparing the last two equations we find $\xi\gtrsim 59\exp\{\pi m_g/g^2 \}$,
which confirms the validity of $\xi\gg 1$ even for $g^2/m_g={\rm O}(1)$.

Numerical checks confirm the computation: the agreement with
the tachyonization estimate is reasonable. For example,
\begin{center}
\begin{tabular}{l|ll}
  $g^2/m_g$ & estimated value of $m_g^2\theta^{\rm crit}$ & numerical value of $m_g^2\theta^{\rm crit}$
  \\ \hline
  $1$ & $0.9\times 10^3$ & $3\times 10^3$
  \\
  $0.33$ & $5\times 10^5$ & $3\times  10^{6}$
  \\
  $0.1$ & $10^{15}$ & $ 10^{16}$
\end{tabular}
\end{center}
For small $g^2/m_g$, one must take extremely large values of $m_g^2\theta$, but the
transition is nevertheless still in the domain of perturbation theory, as
we will discuss in more detail in the next section.

Before we proceed, let us make a few remarks on the $\mu> 1/2$ case.
To evaluate the  ``width'' (in $\mu$) of the
tachyonic regime one can inspect the planar limit
of the full one-loop quantum effects, and notice that
tachyonization is possible whenever
the planar part contributes a term that is greater than the
particle's bare mass
\begin{align}
    \frac{g^2}{8\pi m_g}
    \LL\{
    27\log 3
    -4
    +4{\mu}
    -(1+2\mu)^2
    \log\LL(\frac{2{\mu}+1}{2{\mu} -1}\RR)
    \RR\}
    <
    -1
\end{align}
and \textit{$\xi$ is large enough}. The left hand side is negative only in
a narrow window around $\delta_\mu\doteq \mu-1/2=0$.
For $\delta_\mu>0$ we find that there can be vacuum tachyonization
only for
\begin{align}
  \label{MuWidth}
  \delta_\mu
  <
  \frac{\exp\LL\{\frac{1}{2}-\frac{2\pi m_g}{g^2}\RR\}}{3^{27/4}}
  +{\rm O}(\delta_\mu)^2
  .
\end{align}
For $g^2/m_g=\frac{1}{3}$ this means $\delta_\mu<6.5\times 10^{-12}$, a thin window indeed.

Numerical checks confirm that a  transition to the tachyonic phase,
analogous to the one described above,  is generated for values of
$\mu$ in the range ~\eqref{MuWidth}. We will not discuss this case
in more detail here\footnote{
  An analytic treatment is possible by using, for example,
  \begin{align}
    \int_{-1/2}^{1/2}\dd x\,
    \frac{e^{-\xi\sqrt{x^2+\mu^2-1/4}}}{\sqrt{x^2+\mu^2-1/4}}
    =
    \int_{- 1/ \sqrt{4\mu^2-1} }^{ 1/ \sqrt{4\mu^2-1} }\dd x\,
    \frac{e^{-\xi\sqrt{\mu^2-1/4}\sqrt{x^2+1}}}{\sqrt{x^2+1}}
    \approx
    2K_0(\xi\sqrt{\mu^2-1/4})
    ,
  \end{align}
  where we have stretched the integral's suppressed tails to $\pm\infty$.
}.

Summing up, the picture of the UV/MR mixing is that while in commutative quantum field theories all
information about the bare theory is cancelled by renormalization,
in our case the shifts in the masses are physically observable in the momentum-dependence of the theory\footnote{
  In this sense,  the unexpected appearance of the Euler--Mascheroni constant in the observable
  quantities~\eqref{deltam_mg=0},~\eqref{deltam_g_threshold} and~\eqref{Disper1/2} is a nice signature of the
  ``visibility" of the bare theory after renormalization --- as we remarked above,
  in the commutative theories $\gamma$ (which \textit{e.g.} appears in
  dimensional regularization) always cancels in obervable quantities.
}.
When the mass shifts are negative and sufficiently large, tachyon modes can be generated.

\section{Consistency checks, and the nature of the IR divergences}\label{sec:IRstuff}

As we have seen in the previous sections, tachyonization occurs
because of terms in perturbation theory which are large compared
with the dimensionless expansion parameter $g^2/m_g^2$.
Does this signal a breakdown of perturbation theory?
In this section we address this question by analyzing higher-order corrections, and by
performing some consistency checks on our results.

\subsection{Higher orders in perturbation theory}\label{sec:HigherOrders}

The authors of~\cite{Minwalla:1999px} studied the IR/UV mixing
in the case of the $\lambda \phi^{3\star}$ theory in
$d=6$ and found that the IR/UV effects do not lead one away
from the domain of perturbation theory, as long as
one has ``perturbatively small'' values of $\xi$:
\begin{eqnarray}
  \label{MRS}
  e^{-c/\lambda^2}
  \lesssim
  M^2(p \bullet p)
  \lesssim
  \lambda^2
  ,
\end{eqnarray}
where $M$ is the scalar's mass and $c$ is a constant.
In the previous section, we have found in~\eqref{BigTheta}
that the $\mu\gtrsim 1/2$ tachyon appears for ``nonperturbatively large'' values of $m_g^2\theta$.
An analogue of~\eqref{MRS} for non-commutative YMCS
might mean that the tachyon $\mu\gtrsim 1/2$ is just an artifact of perturbation theory.

Let us review therefore the arguments leading to this relation.
Order by order in perturbation theory, the most divergent
diagram in $d=6$,  $\lambda \phi^{3\star}$ theory is of the following type:
\begin{center}
  \begin{picture}(130,80)(-65,-40)
    \Line(-60,0)(60,0)
    \CCirc(0,0){20}{Black}{White}
    \CCirc(0,14){12}{Black}{White}
    \CCirc(0,23){7}{Black}{White}
  \end{picture}
\end{center}
At order $n$ in perturbation theory, this diagram will get
a quadratic divergence from the ``smallest'' loop in the diagram,
and a logarithmic divergence from each of the $n-1$ bigger loops,
so that its non-planar counterpart will yield an IR-divergent contribution
\[
  \frac{\lambda^{2n}}{p\bullet p}[\log M^2(p\bullet p)]^{n-1}
  .
\]
For this to be of the same magnitude or smaller than the one-loop term,
so that perturbation theory is not invalidated, we must have
\[
  c\frac{\lambda^{2}}{p\bullet p}
  \gtrsim
  \frac{\lambda^{2n}}{p\bullet p}[\log M^2(p\bullet p)]^{n-1}
  ,
\]
which leads directly to the left-hand side of~\eqref{MRS}.
In our language, the right-hand side of that relation is just the request
that the IR-divergent piece overcome the scalar's bare mass term in
the dispersion relation.

Let us apply this same reasoning to our case.
At order $(g^2/m_g)^{n}$, the analogous diagram in our case would be
a nested graph generated by four-gluons vertices:
\begin{center}
  \begin{picture}(120,120)(-60,-25)
    \Gluon(-60,0)(60,0){5.0}{6}
    \GlueArc(0,20)(15,-90,270){5.0}{6}
    \GlueArc(0,50)(8,-90,270){3.5}{6}
    \GlueArc(0,66)(4,-90,270){2.0}{6}
  \end{picture}
\end{center}
In this diagram, however, we only have one (linearly) divergent contribution
from the topmost loop, since the other loops are UV-finite by power counting.
Hence, the condition is
\[
  c\frac{g^{2}}{m_g\sqrt{p\bullet p}}
  \gtrsim
  \frac{g^{2n}}{m_g^n\sqrt{p\bullet p}}
\]
which is satisfied for small enough $g^2/m_g$.
This is clearly due to the fact that topologically massive QED is super-renormalizable
by power counting, and hence we do not expect higher orders in perturbation
theory to invalidate the findings of the previous sections, not even
for what concerns the \mbox{$\mu\gtrsim 1/2$} destabilization and its
exponentially high values of $\theta$.

Even if these large coefficients do not waste the perturbative series,
one might be led to wonder, however, if an improvement in the
approximations could be brought about by studying the full
equation~\eqref{pole1}
\[
  p^2
  =
  \LL[ m_g^{\rm R}(p,\tilde  p)\RR]^2= \frac{m_g^2 \mathcal{Z}_m^2(p,\tilde p)}{\mathcal{Z}_1(p,\tilde p)
  \mathcal{Z}_2(p,\tilde p)}
  ,
\]
and not just its expansion~\eqref{disper1}
\[
  E^2  = |\vec{p}|^2+
  m_g^2\left[1-\left(\frac{g^2}{m_g}\right)\bigg(2\Pi^{\rm o}(p,\tilde
  p)-\Pi^{\rm e}_1(p,\tilde p)- \Pi^{\rm e}_2(p,\tilde p)\bigg)
  \right]_{p^2=m_g^2}
  ,
\]
since the improved equation might (in principle) bring about a
restoration of $E^2>0$.
This problem is analyzed in the next section.

\subsection{The ``improved'' dispersion relation}\label{sec:futility}

Taking a skeptical stance, one might suspect studying the improved version of
equation~\eqref{pole1} to be
non-self-consistent, considering that it implies working at
different orders in the perturbative expansion of the Feynman
diagrams and in the dispersion relations.
However, one should notice that in passing from the
improved equation~\eqref{pole1} to the approximate one~\eqref{disper1},
an expansion in $g^2/m_g$ has been performed, which might
not be meaningful when there are large coefficients in front
of $g^2/m_g$, as in the case of tachyonic momenta.
It is therefore entirely possible for the improved equation~\eqref{pole1}
to give substantially different solutions with respect to~\eqref{disper1},
and this would cast a shadow of doubt on the tachyonic modes found previously.
This section provides a consistency check on the solutions we have found.

To have tachyonic instabilities, by continuity in the squared
momentum $p^2$, there must exist solutions of $p^2=0$ in the full parameter space,
so we need to investigate what is the behavior of the contributions to the polarization tensor for
small values of $\eta^2=p^2/m_g^2$.
This limit competes rather often with that of small $\xi$,  and one must be careful not to mix the two limits.
Remember that these correspond
to the two physically different cases of (respectively)
almost tachyonic renormalized mass $m_g^{\rm R}=0$ for the gauge boson,
and small spatial momentum $|\vec p|$,

Let us first of all investigate the case of $\eta^2\approx 0$ with $\xi$ finite,
which allows to study the onset of tachyonization for arbitrary
values of the spatial momentum $|\vec p|=\xi/(m_g\theta)$.
$\Pi^{\rm o}$ is clearly regular as $\eta\rightarrow 0$,
and an explicit computation shows that so is $\Pi^{\rm e}_1$, thanks to several cancellations.
$\Pi^{\rm e}_2$, on the other hand, has an $1/\eta^2$ divergence,
and this fact must be taken into account when  approximating equation~\eqref{pole1}.
When $\eta^2\approx 0$, $\Pi^{\rm odd}$ and $\Pi^{\rm e}_1$ tend to
well-defined small values (in $g^2/m_g$),
and $\Pi^{\rm e}_{2}\approx {D}/{\eta^2}$ where
\begin{align}
  \begin{split}
    D(g,\mu,\xi)
    \doteq
    \lim_{\eta\rightarrow 0}(\eta^2 \Pi^{\rm e}_2)
    ={}&
    \frac{g^2}{2\pi m_g}
    \LL(\frac{e^{-{\mu}\xi}(1+{\mu}\xi)-e^{-\xi}(1+\xi)}{\xi}\RR)
    .
  \end{split}
\end{align}
The improved pole equation then reads
\begin{align}\label{exact_pole_in_practice}
  \eta^2
  =
  \frac{(1-\Pi^{\rm o})^2}{(1-\Pi^{\rm e}_1)(1-\Pi^{\rm e}_{2})}
  =
  \frac{\LL(1+\Pi^{\rm e}_1-2\Pi^{\rm o}\RR)}{(1-\Pi^{\rm e}_{2})}
  +{\rm O}\LL(\fract{g^4}{m_g^2}\RR),
\end{align}
because $\Pi^{\rm odd}$ and $\Pi^{\rm e}_1$ are always much smaller than
1, at least as long as $g^2/m_g\ll 1$:
the limit of small $\eta^2$ does not interfere with the ``perturbative'' limit here.
Therefore, for $\eta\approx 0$ we may write
  \begin{align}
    \label{FirstApprox}
    \eta^2
    =
    \frac{\eta^2\LL(1+\Pi^{\rm e}_1-2\Pi^{\rm o}\RR)}{\eta^2-D}
    +{\rm O}\LL(\fract{g^4}{m_g^2},\eta^4 \fract{g^4}{m_g^2}\RR)
    ,
  \end{align}
and to have a non-trivial solution with $\eta\approx 0$, we must have $D\approx
-1$. For large $\mu$, we see that $D$ has a minimum for small $\xi$ for which
$D_{\rm min}\approx -\frac{1}{21}{\mu}(g^2/m_g)$. This means that there exist
values of $\eta^2\approx 0$ solving the pole equation for some $\xi$ only if
\begin{align}
  \label{large_mu_futile}
  \mu \frac{g^2}{m_g}\gtrsim 21
  .
\end{align}
Unexpectedly, this is just the perturbative result~\eqref{large_mu_1},
and the improved equation does not give any corrections.

One can also see that, for $\eta^2\approx 0$, the ``exact'' dispersion relation is very similar to the ``approximate'' one.
We are free to say that since $\Pi^{\rm o}(\eta^2=0)$ and  $\Pi^{\rm e}_1(\eta^2=0)$ are much smaller than one,
they do not differ much from their approximate values $\Pi^{\rm o}(\eta^2=1)$ and $\Pi^{\rm e}_1(\eta^2=1)$.
In addition, a numerical computation reveals that
\begin{align}
  \label{numerical_futility}
  \LL|
  D\LL(g,\mu,\xi\RR)
  -\Pi^{\rm e}_2\LL(g,\eta=1,\mu,\xi\RR)
  \RR|
  <
  0.2 \, \frac{g^2}{m_g}
\end{align}
for all $\mu>1/2$, and for all values of $\xi$.
This is not at all evident \textit{a priori}, since $D$ comes from a small $\eta^2$ expansion and $\Pi^{\rm e}_2$ is evaluated
for $\eta^2=1$, and there must be some reason why this is happening. 
Before explaining this, let us have a look at the other case, that is the small-$\xi$ region.

Let us now concentrate on the $\mu\gg 1$ and $\xi\approx 0$ limit.
When $\mu\gg 1$, the fermionic contributions are exponentially suppressed,
and in the $\xi\ll 1$ limit $\Pi^{\rm o}$ is regular, and so is $\Pi^{\rm e}_1$ thanks to
several unexpected cancellations; $\Pi^{\rm e}_2$ only gets contributions from the
gauge sector, and we find
\begin{align}
    \Pi^{\rm e}_{2}
    =
    -\frac{g^2}{2\pi m_g}\left(\frac{1}{\eta^2 \xi}\right)+\frac{g^2}{m_g}{\rm O}(\xi)
    ,
\end{align}
Proceeding as in equation \eqref{FirstApprox}, 
the improved pole equation~\eqref{exact_pole_in_practice} becomes,
thanks to the unexpected $\eta^{-2}$ factor,
\begin{align}
\begin{split}
  \label{disper1_futile}
    \eta^2
    ={}&
    1-2\Pi^{\rm o}(\eta^2)+\Pi^{\rm e}_{1}(\eta^2)
    -\frac{g^2}{2\pi m_g}\left(\frac{1}{\xi}\right)
    +\LL(\fract{g^4}{m_g^2},\fract{g^2}{m_g}\xi\RR)
    =\\={}&
    1-2\Pi^{\rm o}(\eta^2)+\Pi^{\rm e}_{1}(\eta^2)+\Pi^{\rm e}_{2}(\eta^2=1)
    +{\rm O}\LL(\fract{g^4}{m_g^2},\fract{g^2}{m_g}\xi\RR)
    ,
\end{split}
\end{align}
which is roughly equal to the approximate result~\eqref{disper1}.
We stress that we did not use $\eta\ll 1$ in deriving this equation,
but only $\mu\gg 1$ and $\xi\ll 1$.
This finding is just as puzzling as those of~\eqref{large_mu_futile} and~\eqref{numerical_futility},
since it seems that taking $\xi\approx 0$ inside the improved equation
somehow kills all the higher order corrections originating from $\eta^2\ll 1$.

So why are the improved~\eqref{large_mu_futile} and~\eqref{disper1_futile}
and approximate results~\eqref{large_mu_1} and~\eqref{disper1}
approximately equal, that is, why does the (numerical) relation
\eqref{numerical_futility} hold?
The answer is in fact simple, and is related to the nature of the IR divergences, as we anticipated.
Inspecting $\Pi^{\rm e}_2$, we see that all
terms which contain poles in $1/\xi$ also diverge\footnote{
  There are several cancellations at work in $\Pi^{\rm e}_1$, which are not immediately
  evident at first sight.
}
as $1/\eta^2$.
Hence, the two limits $\eta\approx 0$ and $\xi \approx 0$ are connected by the fact that the leading IR divergences have the form
\begin{align}
  \label{this_explains_it_all}
  \Pi^{\rm e}_2(\xi\rightarrow 0)\sim
  \frac{g^2}{m_g}
  \LL(\frac{1}{\eta^2\xi}\RR)
  =
  \frac{g^2}{\theta |\vec p| p^2}
  .
\end{align}
This combination is generated because
\eqref{this_explains_it_all} is
the only combination of the dimensionless quantities
$(g^2/m_g,\mu,\xi,\eta)$ which does not depend on the masses $m_f$ and $m_g$.
It is a well-known fact that also the leading (would-be)
ultraviolet divergences of topologically massive QED are
mass-independent (as some power counting easily shows).

This has been a raher long discussion, so let us sum up the main conclusions of this section.
First of all, the ``coincidence'' that the leading IR divergences have the form \eqref{this_explains_it_all}
ensures that the dispersion relation \eqref{disper1} is indeed a good approximation
of equation \eqref{pole1}: any other pattern of divergences would have wasted the expansion
in powers of $g^2/m_g$.
At the same time, we can take the fact that equation \eqref{this_explains_it_all} is mass-independent 
as an indication that the IR divergences of our theory are a physical phenomenon induced by
the IR/UV mixing, and are not artifacts induced by misplaced
gauge effects \cite{Ruiz:2000hu}, or by a careless truncation of the perturbative series.

\subsection{Off-shell unitarity: intrinsic pinching of the self-energy}\label{sec:Pinch}

In the preceding sections we were able to investigate the
unitarity of the theory only on-shell.
This is because off-shell propagators contain gauge-dependent degrees of freedom (like ghosts),
which hide the unitarity structure of the theory.
To dicuss the theory's unitarity in a more general situation, several different
approaches are available, like employing the background field method \cite{Alvarez-Gaume:2003mb}.

Pinch techniques (PT), in the original formulation by Cornwall
\cite{Cornwall:1981zr,Cornwall:1984eu}, provide a manageable
solution to this problem. These consist in an algorithm that
rearranges the $S$-matrix elements of gauge theories and produces
off-shell proper correlation functions which satisfy the same Ward
identities (WI) as those produced by the classical Lagrangian
\cite{Cornwall:1981zr,Cornwall:1984eu,Haeri:1988af,Papavassiliou:1989zd,Degrassi:1992ue}.
The PT off-shell Green functions, in addition to being gauge
invariant by construction, also satisfy basic theoretical
requirements such as unitarity, analiticity and renormalizability.
They can be also used as the building blocks of gauge-invariant
Schwinger--Dyson equations, which allow to discuss
non-perturbative questions as vacuum stability, dynamical mass
generation \cite{Cornwall:1981zr,Cornwall:1984eu}, and the
behavior of unstable states \cite{Papavassiliou:1995gs,Papavassiliou:1996zn}
in the commutative setup.
An all-orders generalization of the PT has been developed 
both for QCD and for the electroweak sector of the standard
model \cite{Binosi:2002ft,Binosi:2003rr,Binosi:2004qe}.
A comprehensive discussion of this topic is out
of the goals of the present paper (see \cite{Pilaftsis:1996fh} for
an excellent review); here, instead, we shall briefly outline how
these techniques can be applied to the case we are interested in.

In~\cite{Caporaso:2005xf}, the pinch techniques were adapted to
the non-commutative setup in the case of four-dimensional
non-commutative QED. Performing analogous calculations in the
topologically massive theory would be a rather daunting task,
given the number of competing diagrams contributing to the
cancellation. In this paper we shall take a slightly different
approach, using so-called \textit{intrinsic pinch techniques}
\cite{Cornwall:1989gv}.
The intrinsic method essentially consists in showing that all seemingly
unitarity-violating terms in the off-shell propagators cancel when
the equations of motion are imposed. The relation with the method
outlined above is well known: the pinched diagrams which are to be
combined with the ones contributing to the polarization tensor are
all lacking a propagator on an external leg.
This means that all pinching contributions in $\Pi_{\mu\nu}$ must contain an inverse
propagator attached to the index $\mu$ or $\nu$. Similar
considerations apply to the fermion's self-energy, on which we
shall concentrate in a moment.

The intrinsic pinch techniques provide a stronger check of unitarity than simply going on-shell,
because the gauge-dependent terms are dropped before the momentum integrations are carried out.
With respect with the traditional pinch techniques, unfortunately, the intrinsic
approach hides the mechanism by which contributions from different
diagrams conspire to provide off-shell unitarity.
This means that the internal consistency is somewhat less clear,
especially in the prospect of performing a gauge-invariant resummation.
The way in which the spurious threshold cancel is however instructive
and encouraging, as we now show.

Recall that the Majorana self-energy term had the form
\begin{align}
  \begin{split}
    -i\Sigma_{ph}(k)
    =
    2g^2 C \int \frac{\dd^3p}{(2\pi)^3}\; &
    \frac{ \gamma^\mu\LL[\ssh k+\ssh p -m_f\RR]\gamma^\nu
    \LL[p^2{{{g}}}_{\mu\nu}-p_\mu p_\nu -im_g \varepsilon_{\mu\nu\alpha}p^\alpha\RR]}
    {\LL[(k+p)^2-m_f^2\RR]\LL[p^2(p^2-m_g^2)\RR]}
    2\sin^2\LL(\frac{p\vartheta k}{2}\RR)
    ,
  \end{split}
\end{align}
which contains a physical threshold at $p^2=(m_f+m_g)^2$, and an unphysical one at $p^2=m_f^2$.
An ``intrinsic pinch'' consists in identifying all the terms
which contain spurious thresholds, and showing that all these terms contain
an inverse propagator $S^{-1}(q)$ attached to an external leg.
This is equivalent to demanding that all (potentially) unitarity-violating terms cancel
when the equations of motion are imposed.
The integrand can be written as
\begin{align}\label{IntrinsicIntegrand}
    -\frac{ \ssh k+\ssh p +3m_f}
    {\LL[(k+p)^2-m_f^2\RR](p^2-m_g^2)}
    -i m_g
    \frac{ \gamma^\mu S(k+p) \gamma^\nu \varepsilon_{\mu\nu\alpha}p^\alpha}
    {p^2(p^2-m_g^2)}
    -
    \frac{ \ssh p S(k+p)\ssh p}
    {p^2(p^2-m_g^2)}
    ;
\end{align}
using the ``trivial Ward identity''
\begin{align}
  \gamma^\mu S(q) \gamma^\nu \epsilon_{\mu\nu\alpha}p^\alpha
  =
  -i\frac{\LL\{ S^{-1}(q),\ssh p \RR\}}{q^2-m_f^2}
  ,
\end{align}
we rewrite the integrand~\eqref{IntrinsicIntegrand} as
\begin{multline}
    -\frac{ \ssh k+\ssh p +3m_f}
    {\LL[(k+p)^2-m_f^2\RR](p^2-m_g^2)}
    - m_g
    \frac{\LL\{ S^{-1}(k)+\ssh p,\ssh p \RR\}}{\LL[(p+k)^2-m_f^2\RR] p^2(p^2-m_g^2)}
    +\\
    -
    \frac{ \ssh p S(k+p)  \LL[ S^{-1}(k+p)-S^{-1}(k) \RR]}
    {p^2(p^2-m_g^2)}
    .
\end{multline}
By definition, all $S^{-1}(k)$ terms are ``pinching'', since they vanish when
the equations of motion are imposed.
In addition, all such terms contain spurious thresholds, as expected if unitarity holds.
The pinched self-energy becomes
\begin{align}
  \begin{split}
    -i\Sigma_{ph}(k)
    ={}&
    -2g^2 C \int \frac{\dd^3p}{(2\pi)^3}\;
    \frac{ \ssh k+\ssh p +3m_f+2m_g}{\LL[(k+p)^2-m_f^2\RR](p^2-m_g^2)}
    \LL(1-\cos p\vartheta k \RR)
    ;
  \end{split}
\end{align}
notice that we did not need to perform the momentum integration to define this pinched quantity.
Now we can introduce an $x$ parameter and perform the integral,
getting (in units of $m_g$)
\begin{align}
  \begin{split}
    -i\frac{\Sigma_{ph}(k)}{m_g}
    ={}&
    -\frac{ig^2}{4\pi m_g}
    C \int \dd x\;
    \frac{(1-x)\ssh \eta +3\mu+2}{\sqrt{x\mu^2-x(1-x)\eta^2+(1-x)}}
    \LL(
    1-e^{-\xi\sqrt{-x(1-x)\eta^2+x\mu^2+(1-x)}}
    \RR)
    .
  \end{split}
\end{align}
Comparing this expression with the unpinched one (see section~\ref{sec:majo_self}),
we see that the unphysical threshold $p^2=m_f^2$ has disappeared, while the physical one
at $p^2=(m_f+m_g)^2$ survives.

A consistency check of this approach is obtained by verifying that
on shell the propagator is unchanged, as one would expect if only
gauge artefacts are removed by the procedure.
On shell $(\ssh k=-\mu)$ one easily sees that indeed the pinched dispersion relation is
identical to the unpinched one~\eqref{deltam_f_soft}, as expected.

We have performed an analogous, quite lengthy and rather uninstructive computation,
veryfying that the same mechanism holds for the gauge boson's propagator:
all the potentially unitarity-violating terms corresponding to the thresholds
at $p^2=0$ and $p^2=m_g^2$ vanish with the intrinsic pinch techniques,
and the pinched propagator is identical to the standard one in the on-shell limit.

\section{Conclusions and outlook}\label{sec:conclusions}

Our findings suggest that non-commutative, softly supersymmetric Yang--Mills--Chern--Simons is a well-defined theory, 
possessing a perturbatively stable vacuum in a large part 
of its parameters space of couplings and masses $(g,m_g,m_f,\theta)$.

The perturbative vacuum becomes unstable in two different regions, 
corresponding to $m_f\gg m_g$ and ($m_f\gtrsim m_g/2$, $m_g^2\theta\gtrsim 2\,e^{3+\pi m_g/g^2}$), 
as we have checked at one-loop, and verified against higher-order corrections. 
The physics at work in these two regions appear to be very different: in the case of large $m_f$, 
tachyonic instabilities are triggered by an imperfect cancellation of the IR/UV mixing phenomenon. 
In the other region, $m_f\gtrsim m_g/2$, we observe a peculiar ``MR/UV'' mixing, by which a divergence 
in the mass renormalization of the commutative theory is transported in the ultraviolet regime of the non-commutative model.

At the moment, the nature of the new vacuum is a matter of speculation~\cite{Barbon:2001dw}.
In our model, the tachyonic mode is characterized by a well-defined, finite momentum,
as can be seen in figures~\ref{fig:DispersionRelationDifferentMu} and~\ref{SoftMassShiftNearThreshold2}.
For this reason, one may conjecture that this instability
will lead the system to a sort of stripe phase analogous to that proposed by Gubser and
Sondhi~\cite{Gubser:2000cd,Castorina:2003zv,Mandanici:2003vt}
for the $\phi^{4\star}$ theory.
This possibility is however rather problematic: a non-translationally invariant vacuum
would bring about a dynamical breaking of the gauge invariance,
endangering the consistency of the entire theory.
Recall that, for a non-commutative gauge theory, spacetime translations are in fact
a subset of the gauge transformations.

A less speculative (but nevertheless very intriguing) point of view
is to suppose that the tachyonic mode will drive the
Yang--Mills--Chern--Simons system through a phase transition
similar to the one speculated by
Cornwall~\cite{Cornwall:1981zr,Cornwall:1984eu} for the
non-Abelian model, in the commutative case.
The fate of the perturbative vacuum should be discussed, of
course, at the non-perturbative level 
\cite{Eguchi:1982nm,Gonzalez-Arroyo:1982hz,Bhanot:1982sh,Ambjorn:1999ts,Ambjorn:2000nb,Griguolo:2003kq}.
From this point of view, the lattice formulation is particularly
promising 
\cite{Ambjorn:2000cs,Nishimura:2001dq,Nishimura:2002hw,Bietenholz:2003hx,Bietenholz:2004xs,Bietenholz:2004xv,Bietenholz:2005iz,Bietenholz:2006cz};
in addition, the search for exact solutions of non-commutative quantum field
theories can shed some light on some aspects of these theories, like
the IR/UV-related phenomena \cite{Langmann:2002cc,Langmann:2003cg,Langmann:2003if}.

Another course, which was outlined in section~\ref{sec:Pinch} and discussed in some detail in
\cite{Caporaso:2005xf}, would be to employ a gauge-invariant
resummation formalism to study the fate of the perturbative
vacuum. The next natural step in this direction is to compute the
pinched three-vertex in non-commutative QED and YMCS.

\section*{Acknowledgements}

We would like to thank L. Griguolo and D. Seminara for useful discussions
during the making of this paper. The work of N.~C. was partially supported by a
fellowship of the ``Angelo della Riccia'' foundation.

\appendix

\section{Conventions and notations}

We take the metric in flat $2+1$ Minkowski spacetime to be
\begin{align}
    {{{g}}}^{\mu\nu}&={\rm diag}(+1,-1,-1)
    &\mu=0,1,2
\end{align}
and for the $\varepsilon^{\mu\nu\varrho}$ we take
$\varepsilon^{012}=\varepsilon_{012}=1$.

Passing from Minkowski to Euclidean space and back is
a bit tricky when the tensor $\theta_{\mu\nu}$ is around.
To make an analytic continuation in the non-commutative case, one needs to define a
Wick rotation for the matrix $\theta_{\mu\nu}$ that preserves the
Moyal phases, so that integrals become convergent:
this is given by
$    \theta_M^{0i} = -i\theta_E^{0i}$.
Bullets are defined in the spacelike and timelike cases as
\begin{align}
  \label{pBulletp}
    (p\bullet p)_M =
    \begin{cases}
    \theta^2(p_0^2-p_1^2)&\rm{mixed, or}
    \\
    \theta^2(p_1^2+p_2^2)&\rm{spacelike}.
    \end{cases}
\end{align}
The main equation is rather counter-intuitive:
\begin{align}
    +\tilde p_M^2 \equiv  - (p\bullet p)_M,
    \qquad
    \stackrel{\rm Wick}\longleftrightarrow
    \qquad
    - \tilde p_E^2 \equiv  - (p\bullet p)_E,
\end{align}
because it implies that $(p\bullet p)$ is invariant, but its name $(\tilde
p^2)$ changes sign. Therefore,
\begin{align}
    (p\bullet p)_E =
    \begin{cases}
    \theta^2(p_0^2+p_1^2)&\rm{mixed},
    \\
    \theta^2(p_1^2+p_2^2)&\rm{spacelike},
    \end{cases}
\end{align}
and one sees immediately that, in Euclidean space, $(p\bullet p)$
is positive-definite --- as it should be.

\subsection{Feynman rules}\label{sec:FeynmanRulez}

The path integral approach to the quantization of gauge theories
generalizes easily to the non-commutative setup also in the presence
of a Chern--Simons term.
The propagators and vertices are
\begin{itemize}
  \item Gluon propagator:
  \begin{center}
  \begin{picture}(300,40)
    \Gluon(5,20)(90,20){4}{10} \ArrowLine(40,10)(60,10)
    \Text(18,32)[]{$\mu$}\Text(92,32)[]{$\nu$}\Text(55,32)[]{$p$}
    \Text(220,20)[]{
    $\displaystyle{\frac{i}{p^2(p^2-m_g^2)}\left(p^2{{{g}}}_{\mu\nu}-p_\mu p_\nu+im_g\epsilon_{\mu\nu\alpha}p^\alpha\right)+\xi \frac{p_\mu p_\nu}{p^4}}$}
  \end{picture}
  \end{center}
  \item Three-gluon vertex
  \begin{center}
    \begin{picture}(400,100)
    \Gluon(50,100)(50,50){4}{6} \ArrowLine(58,85)(58,65)
    \ArrowLine(86,24)(70,40) \ArrowLine(15,24)(31,40)
    \Gluon(50,50)(10,10){4}{6}\Gluon(50,50)(90,10){4}{6}
    \Text(5,20)[]{$\mu$}\Text(28,12)[]{$q$}
    \Text(95,20)[]{$\nu$}\Text(75,12)[]{$r$}
    \Text(60,90)[]{$\lambda$}\Text(40,90)[]{$p$}
    \Text(200,60)[]{$\mathcal{V}_3^{\lambda\mu\nu}(p,q,r)=(2\pi)^3\delta(p+q+r)2ig\sin\left(\frac{p\wedge q}{2}\right)$}
    \Text(260,40)[]{$(im_g\epsilon_{\lambda\mu\nu}-{{{g}}}_{\lambda\mu}(p-q)_\nu-{{{g}}}_{\mu\nu}(q-r)_\lambda-{{{g}}}_{\nu\lambda}(r-p)_\mu)\doteq$}
    \Text(245,20)[]{$\doteq(2\pi)^3\delta(p+q+r)2ig\sin\left(\frac{p\wedge q}{2}\right)T_3^{\lambda\mu\nu}$}
    \end{picture}
  \end{center}
  \item Four-gluon vertex
  \begin{center}
    \begin{picture}(400,100)
    \Gluon(50,50)(90,90){4}{6}
    \Gluon(50,50)(90,90){4}{6}\Gluon(50,50)(10,90){4}{6}
    \ArrowLine(19,70)(35,54)
    \ArrowLine(86,24)(70,40)
    \ArrowLine(15,24)(31,40)\ArrowLine(80,70)(64,54)
    \Gluon(50,50)(10,10){4}{6}\Gluon(50,50)(90,10){4}{6}
    \Text(5,20)[]{$\nu$}\Text(28,12)[]{$q$}
    \Text(95,20)[]{$\alpha$}\Text(75,12)[]{$r$}
    \Text(23,90)[]{$\mu$}\Text(5,83)[]{$p$}\Text(78,90)[]{$s$}\Text(94,83)[]{$\beta$}
    \Text(250,80)[]{$\mathcal{V}_4^{\mu\nu\alpha\beta}(p+q+r+s)=(2\pi)^3\delta(p+q+r+s)(-4g^2)$}
    \Text(260,60)[]{$\big[\left(\sin\frac{p\wedge q}{2}\sin\frac{r\wedge s}{2}({{{g}}}^{\mu\alpha}{{{g}}}^{\nu\beta}-{{{g}}}^{\mu\beta}{{{g}}}^{\nu\alpha}\right)$}
    \Text(250,40)[]{$+\left(\sin\frac{p\wedge r}{2}\sin\frac{q\wedge s}{2}({{{g}}}^{\mu\nu}{{{g}}}^{\alpha\beta}-{{{g}}}^{\mu\beta}{{{g}}}^{\nu\alpha}\right)$}
    \Text(250,20)[]{$+\left(\sin\frac{q\wedge r}{2}\sin\frac{p\wedge s}{2} ({{{g}}}^{\mu\nu}{{{g}}}^{\alpha\beta}-{{{g}}}^{\nu\beta}{{{g}}}^{\mu\alpha}\right)\big]$}
    \end{picture}
  \end{center}

\end{itemize}
The Feynman rules for the ghosts and fermions are unaffected
by the Chern--Simons term: see for example \cite{Caporaso:2005xf}.

\subsection{Useful integrals}
\label{sec:IntegralsInMinkowski}

In the three-dimensional case, the non-planar integrals \cite{Brandt:2001ud} give rise
to half-integer Bessel functions
\begin{align}\label{nonplanar_integrals}
  \begin{split}
    \int \frac{\dd^3p}{(2\pi)^3} \frac{e^{ip\theta k}}{[p^2-M^2]^2}
    =
    + \frac{i\sqrt{2}}{8\pi^{3/2}}{}&
    \LL(\frac{\sqrt{k\bullet k}}{M}\RR)^{1/2} K_{1/2}\LL(M\sqrt{k\bullet k}\RR)
    \\
    \int \frac{\dd^3p}{(2\pi)^3} \frac{p_\mu p_\nu\; e^{ip\theta k}}{[p^2-M^2]^2}
    =
    -\frac{i\sqrt{2}}{8\pi^{3/2}}{}&
    \Bigg[
      {{{g}}}_{\mu\nu}\LL(\frac{M}{\sqrt{k\bullet k}}\RR)^{1/2} K_{1/2}\LL( M\sqrt{k\bullet k}\RR)
      +\\&
      +(\theta k)_\mu(\theta k)_\nu\LL(\frac{M}{\sqrt{k\bullet k}}\RR)^{3/2} K_{3/2}\LL( M\sqrt{k\bullet k}
                   \RR)
    \Bigg]
    \\
    \int \frac{\dd^3p}{(2\pi)^3} \frac{p^2\;e^{ip\theta k}}{[p^2-M^2]^2}
    =
    -\frac{i\sqrt{2}}{8\pi^{3/2}}{}&
    \Bigg[
      3\LL(\frac{M}{\sqrt{k\bullet k}}\RR)^{1/2} K_{1/2}\LL( M\sqrt{k\bullet k}\RR)
      +\\&
      +(\theta k)^2\LL(\frac{M}{\sqrt{k\bullet k}}\RR)^{3/2} K_{3/2}\LL( M\sqrt{k\bullet k}
                   \RR)
    \Bigg]
     ,
  \end{split}
\end{align}
where we have set  $z \doteq M\sqrt{k\bullet k}$
and $\frac{dz}{d(\theta k)_\mu}
    =
    -\frac{M}{\sqrt{k\bullet k}}(\theta k)^\mu$.
There are some useful recurrence relations:
\begin{align}
    \label{besselk_derivative}
    K_{n+1}-K_{n-1} ={}& -2\frac{\dd K_n}{\dd x}
    ,
\\
  \label{besselk_three}
    zK_{n-1}(z)-zK_{n+1}(z) ={}& -2nK_{n}(z)
    ,
\end{align}
which we have employed extensively to get a simpler expression of the
polarization tensor $\Pi_{\mu\nu}$.

\bibliography{nc_ym_cs_b}

\end{document}